\documentclass[9pt,twoside]{pnas-new}

\templatetype{pnasmathematics} 

\setboolean{displaywatermark}{false}
\usepackage{dsfont} 

\usepackage{mathtools} 

\usepackage{enumitem}

\usepackage{amsthm}
\newtheorem{theorem}{Theorem}

\usepackage{multirow}
\usepackage{booktabs}
\usepackage{graphicx}
\usepackage{caption}
\usepackage{subcaption}
\usepackage[outdir=./]{epstopdf}
\usepackage{siunitx}
\usepackage[ruled,vlined]{algorithm2e}
\newcommand{\timeValue}{t} 

\newcommand{\powerInequality}{\mathcal{I}} 
\newcommand{\calBlue}{\mathcal{B}} 
\newcommand{\calRed}{\mathcal{R}} 
\newcommand{\graph}{G} 
\newcommand{\nodeSet}{V} 
\newcommand{\edgeSet}{E} 

\newcommand{\redBirthProb}{r} 

\newcommand{\probEventOne }{p} 
\newcommand{\probEventTwo }{q} 

\newcommand{\deltaboth}{\delta} 
\newcommand{\deltaIn}{\deltaboth_{i}} 
\newcommand{\deltaOut}{\deltaboth_{o}} 

\newcommand{\homophilyRedAnyEvent}{\rho_{\calRed}} 
\newcommand{\homophilyBlueAnyEvent}{\rho_{\calBlue}}  

\newcommand{\homophilyRedEventOne}{\homophilyRedAnyEvent^{(1)}} 
\newcommand{\homophilyBlueEventOne}{\homophilyBlueAnyEvent^{(1)}}  

\newcommand{\homophilyRedEventTwo}{\homophilyRedAnyEvent^{(2)}} 
\newcommand{\homophilyBlueEventTwo}{\homophilyBlueAnyEvent^{(2)}}  

\newcommand{\homophilyRedEventThree}{\homophilyRedAnyEvent^{(3)}} 
\newcommand{\homophilyBlueEventThree}{\homophilyBlueAnyEvent^{(3)}}  

\newcommand{\matrixAnyEvent}{E} 
\newcommand{\matrixEventOne}{\matrixAnyEvent_{1}} 
\newcommand{\matrixEventTwo}{\matrixAnyEvent_{2}} 
\newcommand{\matrixEventThree}{\matrixAnyEvent_{3}} 

\newcommand{\inDegree}{d_{i}} 
\newcommand{\outDegree}{d_{o}} 
\newcommand{\degree}{d} 

\newcommand{\thetaIn}{\theta^{(i)}} 
\newcommand{\thetaOut}{\theta^{(o)}} 
\newcommand{\thetaVec}{\theta} 

\newcommand{\nonLinFunction}{F} 

\newcommand{\groupSize}{n} 

\newcommand{\poneBB}{{p}^{(1)}_{\calBlue_{\mathrm{new}}\rightarrow\calBlue_{\mathrm{old}}}} 
\newcommand{\poneBR}{{p}^{(1)}_{\calBlue_{\mathrm{new}}\rightarrow\calRed_{\mathrm{old}}}} 
\newcommand{\poneRR}{{p}^{(1)}_{\calRed_{\mathrm{new}}\rightarrow\calRed_{\mathrm{old}}}} 
\newcommand{\poneRB}{{p}^{(1)}_{\calRed_{\mathrm{new}}\rightarrow\calBlue_{\mathrm{old}}}} 

\newcommand{\ptwoBB}{{p}^{(2)}_{\calBlue_{\mathrm{new}}\leftarrow\calBlue_{\mathrm{old}}}} 
\newcommand{\ptwoBR}{{p}^{(2)}_{\calBlue_{\mathrm{new}}\leftarrow\calRed_{\mathrm{old}}}} 
\newcommand{\ptwoRR}{{p}^{(2)}_{\calRed_{\mathrm{new}}\leftarrow\calRed_{\mathrm{old}}}} 
\newcommand{\ptwoRB}{{p}^{(2)}_{\calRed_{\mathrm{new}}\leftarrow\calBlue_{\mathrm{old}}}} 

\newcommand{\pthreeBB}{{p}^{(3)}_{\calBlue_{\mathrm{old}}\rightarrow \calBlue_{\mathrm{old}}}} 
\newcommand{\pthreeBR}{{p}^{(3)}_{\calBlue_{\mathrm{old}}\rightarrow \calRed_{\mathrm{old}}}} 
\newcommand{\pthreeRR}{{p}^{(3)}_{\calRed_{\mathrm{old}}\rightarrow \calRed_{\mathrm{old}}}} 
\newcommand{\pthreeRB}{{p}^{(3)}_{\calRed_{\mathrm{old}}\rightarrow\calBlue_{\mathrm{old}}}} 

\newcommand{\barponeBR}{\bar{p}^{(1)}_{\calBlue_{\mathrm{new}}\rightarrow\calRed_{\mathrm{old}}}} 
\newcommand{\barponeRR}{\bar{p}^{(1)}_{\calRed_{\mathrm{new}}\rightarrow\calRed_{\mathrm{old}}}}

\newcommand{\barptwoBR}{\bar{p}^{(2)}_{\calBlue_{\mathrm{new}}\leftarrow\calRed_{\mathrm{old}}}} 
\newcommand{\barptwoRR}{\bar{p}^{(2)}_{\calRed_{\mathrm{new}}\leftarrow\calRed_{\mathrm{old}}}}

\newcommand{\barpthreeBR}{\bar{p}^{(3)}_{\calBlue_{\mathrm{old}}\rightarrow \calRed_{\mathrm{old}}}} 
\newcommand{\barpthreeRR}{\bar{p}^{(3)}_{\calRed_{\mathrm{old}}\rightarrow \calRed_{\mathrm{old}}}} 
\newcommand{\barpthreeRB}{\bar{p}^{(3)}_{\calRed_{\mathrm{old}}\rightarrow\calBlue_{\mathrm{old}}}} 

\title{Emergence of Structural Inequalities of Scientific Impact}
\title{Emergence of Structural Inequalities in Scientific Citation Networks}
\author[a,c]{Buddhika Nettasinghe}
\author[b]{Nazanin Alipourfard} 
\author[a]{Vikram Krishnamurthy}
\author[b]{Kristina Lerman}

\affil[a]{Cornell University}
\affil[b]{USC Information Sciences Institute}

\leadauthor{Nettasinghe} 

\correspondingauthor{\textsuperscript{2}To whom correspondence should be addressed. E-mail: dwn26\@cornell.edu}


\begin{abstract}
Structural inequalities persist in society, conferring systematic advantages to some people at the expense of others, for example, by giving them substantially more influence and opportunities. Using bibliometric data about authors of scientific publications, we identify two types of structural inequalities in scientific citations. First, female authors, who represent a minority of researchers, receive less recognition for their work (through citations) relative to male authors; second, authors affiliated with top-ranked institutions, who are also a minority, receive substantially more recognition compared to other authors. We  present a model for the growth of directed citation networks and show that citations disparities arise from individual preferences to cite authors from the same group (homophily), highly cited or active authors (preferential attachment), as well as the size of the group and how frequently new authors join. We analyze the model and show that its predictions align well with real-world observations. Our theoretical and empirical analysis also suggests potential strategies to mitigate structural inequalities in science. In particular, we find that merely increasing the minority group size does little to narrow the disparities. Instead, reducing the homophily of each group, frequently adding new authors to a research field while providing them an accessible platform among existing, established authors, together with balanced group sizes can have the largest impact on reducing inequality. 
Our work highlights additional complexities of mitigating structural disparities stemming from asymmetric relations (e.g., directed citations) compared to symmetric relations (e.g., collaborations). 
\end{abstract}

\begin{document}

\maketitle
\thispagestyle{firststyle}
\ifthenelse{\boolean{shortarticle}}{\ifthenelse{\boolean{singlecolumn}}{\abscontentformatted}{\abscontent}}{}

\dropcap{G}rowing concerns about social justice have brought renewed scrutiny to the problem of structural inequalities. Such inequalities concentrate power in the hands of certain groups based on their gender, race, or socio-economic status, by prioritizing their access to resources and social capital, while limiting opportunities for others. Arguably the best known of these is economic inequality, in which a small minority controls the disproportionate share of income and wealth. Economic inequality is detrimental to social welfare and has been associated with adverse societal outcomes, including reduced well-being~\cite{payne2017broken} and increased mortality~\cite{case2020deaths}, crime, and other social problems~\cite{wilkinson2009income}. 

Structural inequalities are also common in science. As a result of the gender gap in faculty hiring, women remain a small minority in many fields: they represent 15\% of tenure track faculty in Computer Science~\cite{way2016gender},  23\% (resp. 10\%) of assistant (resp. full) professors in Physics~\cite{porter2019women}, and 31\% (resp. 15\%) of assistant (resp. full) professors in Economics~\cite{chen2016female}. Although women publish at the same rate as men, their papers appear in less prestigious journals~\cite{ross2020leaky} and tend to receive fewer citations~\cite{dion2018gendered,fulvio2020gender}. 
Academic prestige presents another source of disparities in faculty hiring, where the ranking of a researcher's doctoral degree institution determines their academic placement and career opportunities~\cite{burris2004academic,bedeian2010doctoral}. According to one study of faculty hiring, the top-ranked 15\% of computer science departments produced 68\% of their own faculty, hiring less than 10\% from outside of the top-50 departments~\cite{clauset2015systematic}. The disparity is not driven by competition in the job market for best candidates. Instead, the benefits of early career placement, rather than inherent merit, serve to lock in advantages of doctoral training prestige, thereby facilitating future success~\cite{way2019productivity}. 

In this paper, we demonstrate that structural inequalities extend beyond hiring to affect the recognition scientists receive from their peers, as measured by the number of citations to their papers. Citations provide the basis for evaluating the research impact of scientists, a metric widely used in academia to decide who to promote, invite as a keynote speaker, award tenure or a prize. Citations also help the community and funding agencies identify important research topics.
We demonstrate the existence of a gender gap and a prestige gap in citations that make one group receive substantially less recognition for their work than the other. Specifically, women, who are a minority in science, receive less recognition than men. However, researchers from top-ranked institutions, who also form a minority, receive far more recognition than others; moreover, the smaller and more prestigious the group of researchers from top-ranked institutions, the larger the citations inequality. These two examples show that disparities in representation alone do  not explain structural inequalities in citations.  

To elucidate the origins of these structural inequalities, we present a dynamic model of the growth of directed citations networks, which shows that  inequalities arise as a consequence of biases in individual preferences of authors to cite similar (homophily property of social interactions) and well-connected or active (preferential attachment) authors, as well as the size of the minority group and how quickly new authors are integrated into the citation network. The model is simple enough to be analytically tractable, yet rich enough to capture key elements of real-world dynamics.  Our theoretical analysis and empirical validation reveal a rich array of phenomena 
as well as strategies for reducing citation disparity. 

Researchers have recently begun to study  disparities from the perspective of network science. Studies have shown that the structure of social interactions can systematically disadvantage members of the minority group by limiting their visibility~\cite{lee2019homophily}, relative ranking~\cite{karimi2018homophily}, economic opportunities~\cite{jackson2021inequality} and the number of social  connections they accrue~\cite{avin2015homophily,avin2017modeling}. These studies have focused on undirected networks representing professional relationships among scientists, such as  who works with whom (collaboration networks) or who knows whom (affiliation networks). In contrast to undirected networks where social links are symmetric, we focus on directed citation networks with asymmetric links. 
We use large bibliographic databases containing information about scientific publications and publications they cite, to construct citation networks of authors. A directed edge in the author-citation network means that one author cites the papers of the other author in their own works, but not necessarily the other way around. We enrich these networks with additional information about authors and their affiliations. The scale and scope of the data allow for longitudinal study of disparities and comparison across academic disciplines.

Reducing structural inequalities is critical to broadening participation in science, which is itself necessary not only to promote equity, but also to spur innovation on which our economic prosperity depends. Our analysis shows that merely increasing the size of the minority group, e.g., through hiring policies or affirmative action, does little to change inequality. Instead, our work suggests the need to change the culture of scientific recognition by changing individual citation preferences and inclusion of new authors. 
Our work provides a basis for metrics that scientific publishers and academic search engines can use to audit the content they produce for bias, an important step toward reducing disparities in science.

\section*{Results}
\subsection*{Power-Inequality in Author-Citation Networks}

\begin{figure}
        \centering
        \begin{subfigure}[b]{0.45\textwidth}
            \centering
            \includegraphics[width=\textwidth]{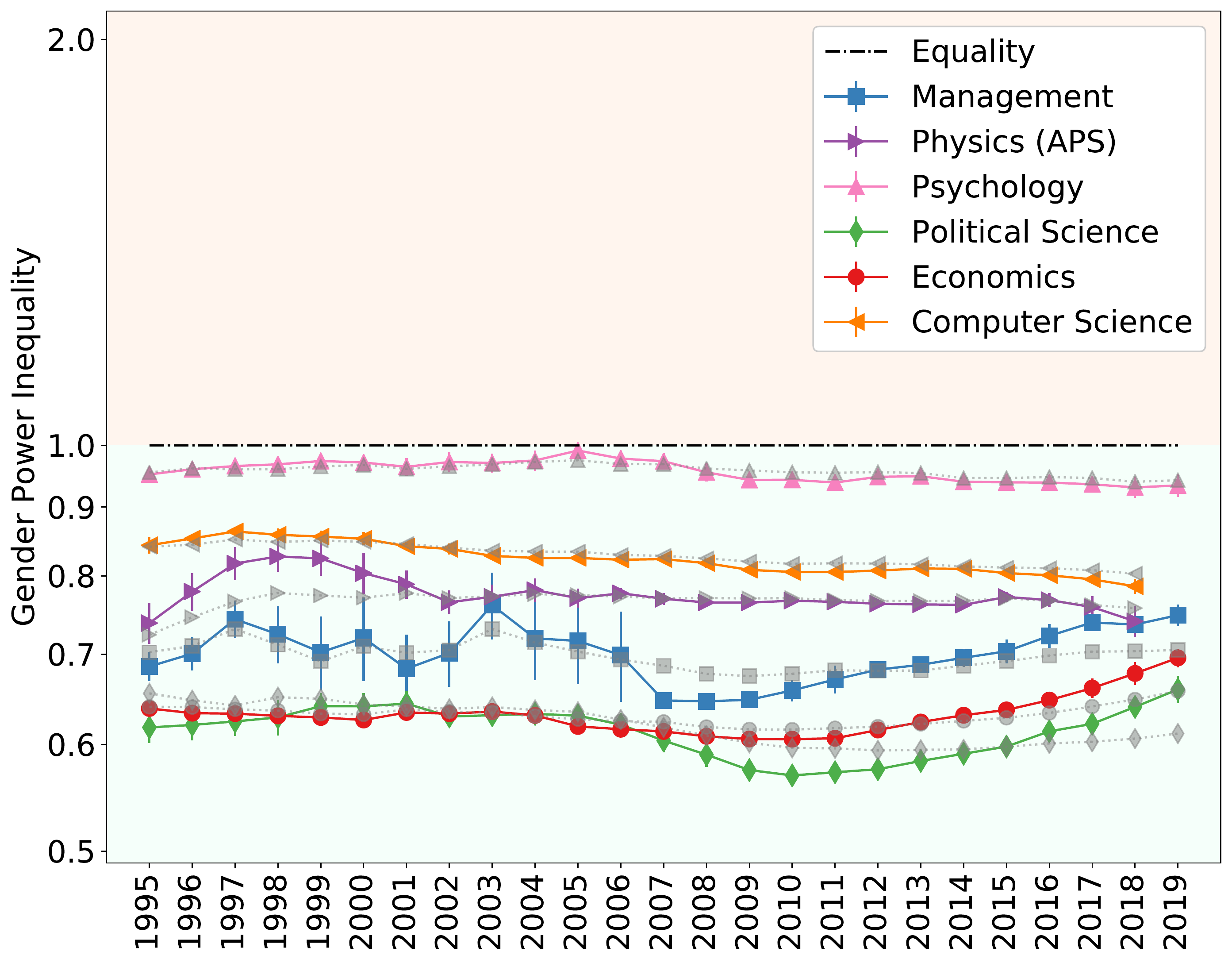}
            \caption[]%
            {{\small Gender Power-inequality}}
            \label{fig:gender_power}
        \end{subfigure} \hspace{0.5cm}
        \begin{subfigure}[b]{0.45\textwidth}
            \centering
            \includegraphics[width=\textwidth]{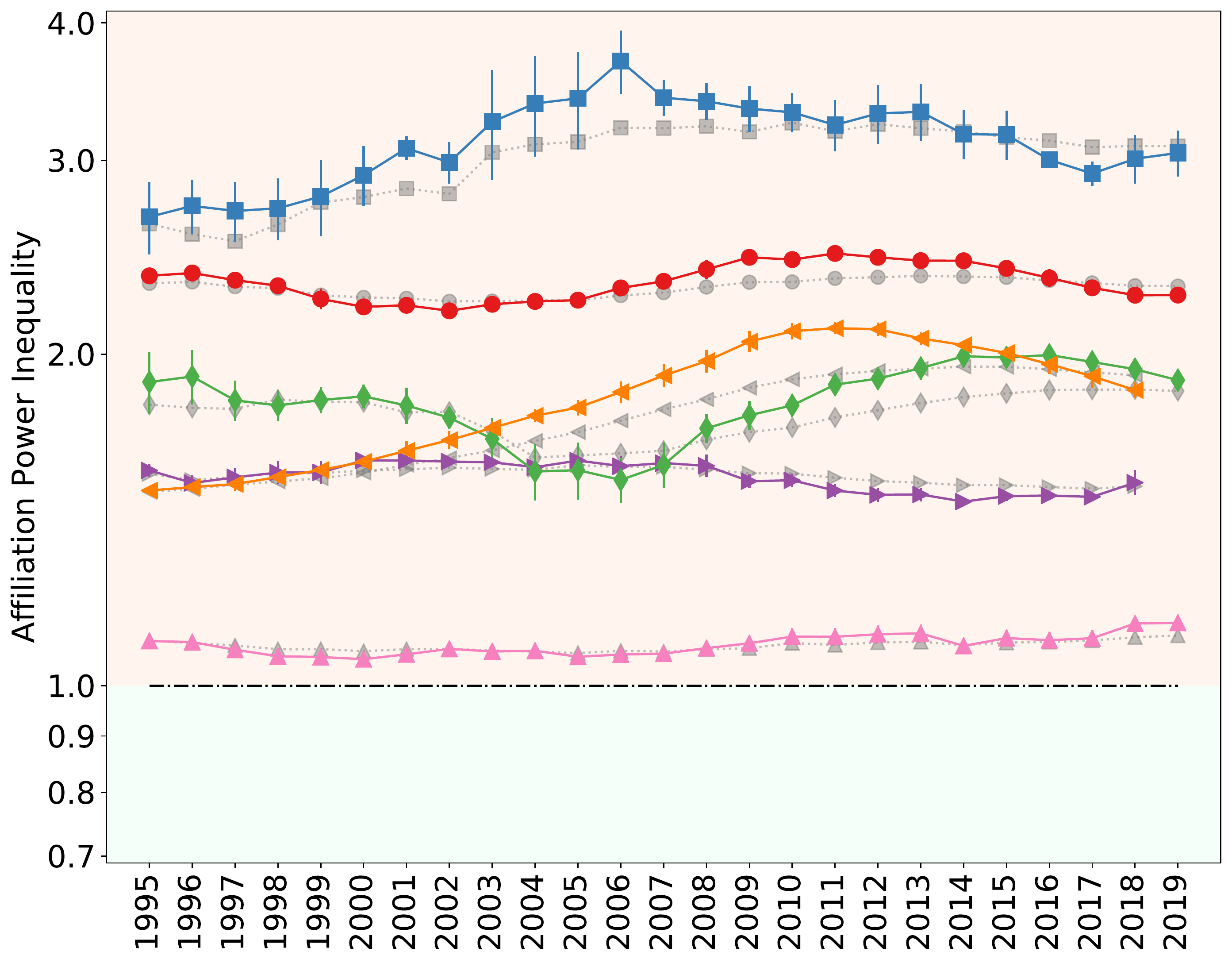}
            \caption[]%
            {{\small Affiliation Power-inequality}}
            \label{fig:aff_power}
        \end{subfigure}
        \begin{subfigure}[b]{0.45\textwidth}
            \includegraphics[width=\textwidth]{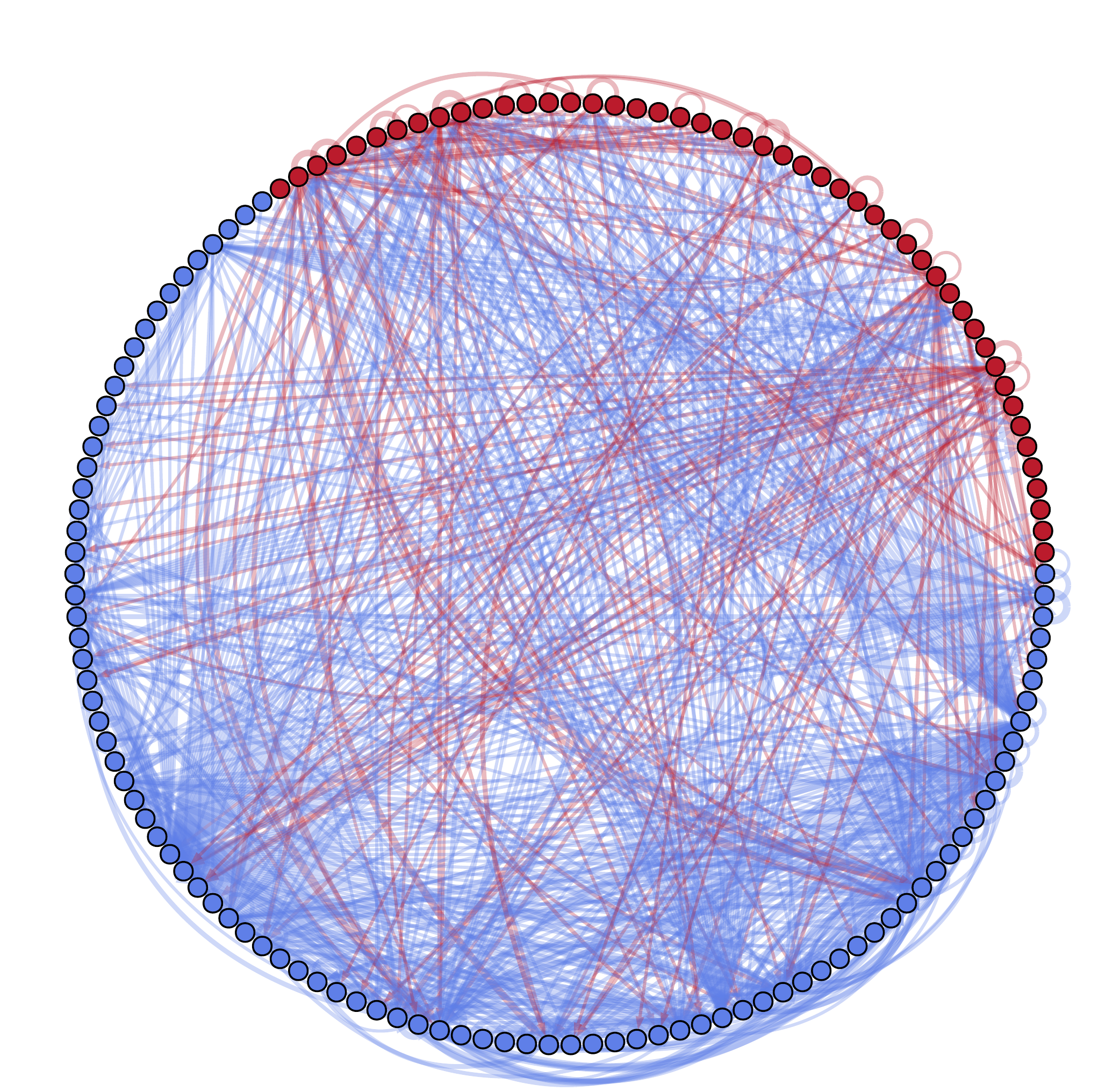}
            \caption[]%
            {{\small A subgraph of gender-partitioned network}}
            \label{fig:aff_power}
        \end{subfigure}
        \hspace{0.75cm} \begin{subfigure}[b]{0.45\textwidth}
            \centering
            \includegraphics[width=\textwidth]{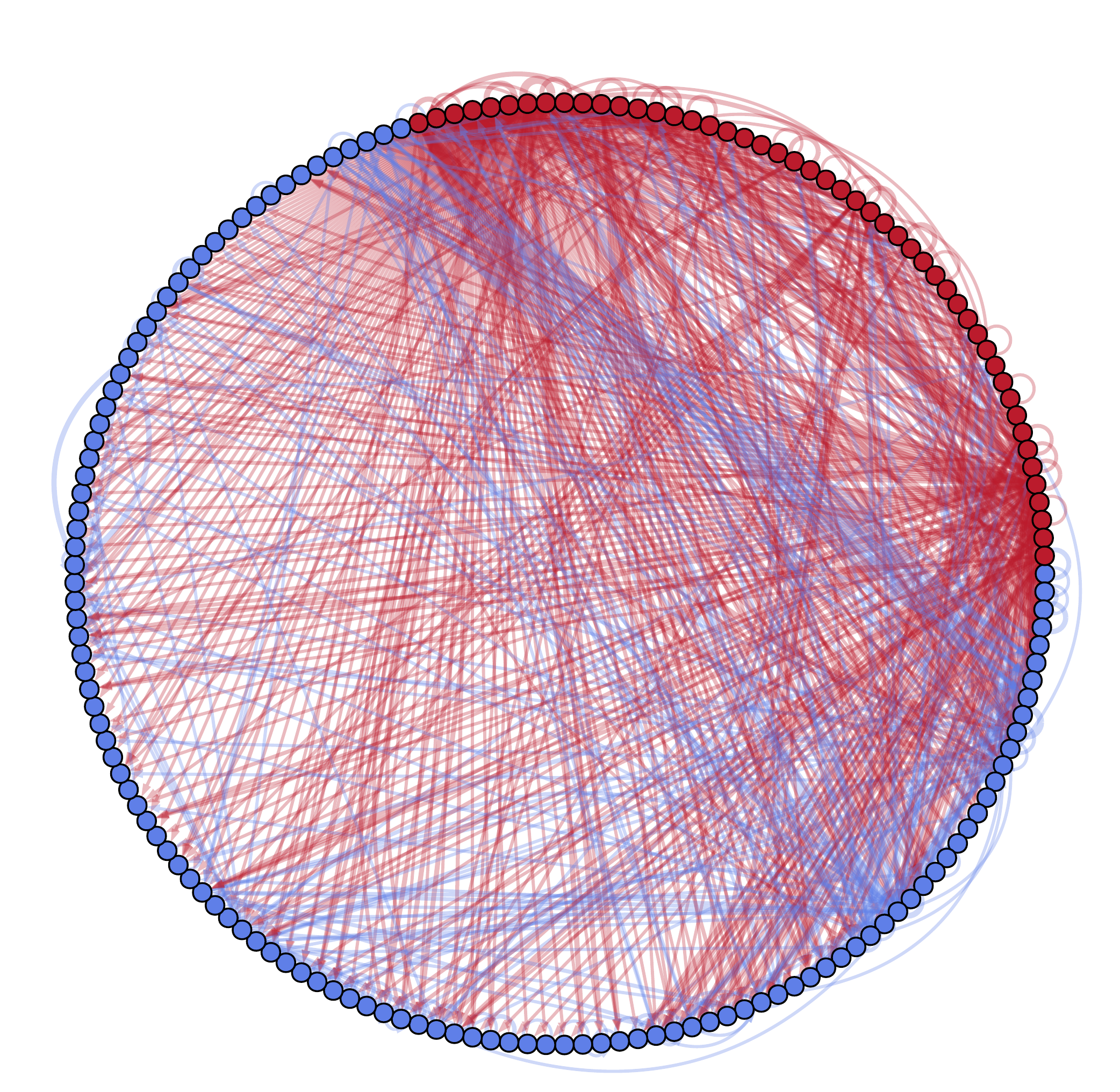}
            \caption[]%
            {{\small A subgraph of the affiliation-partitioned network}}
            \label{fig:aff_power}
        \end{subfigure}
        \caption{
        Power-inequality  over time for six fields of study in \textbf{(a)} gender-partitioned and \textbf{(b)} affiliation-partitioned networks. The plotted values indicate the average of power-inequality, defined in Eq.~(\ref{eq:power_inequality}), over a sliding window of four years, and confidence intervals indicate the standard error. The gray lines show the cumulative power inequality over the years. In \textbf{(a)}, all power-inequality values are below 1, suggesting that female authors have less power than male authors. {Psychology} is the closest field to gender parity. {Economics}, {Management} and {Political Science} have steadily been moving towards gender parity
        after 2010, while {Computer Science} has a slightly decreasing trend over time. In \textbf{(b)}, the minority (red) group represents authors affiliated with the top-100 institutions. Power-inequality values are above 1, suggesting that authors from top-ranked institutions have more power than other authors even though they are the minority. {Psychology} again is the field with values closer to 1 and therefore is the most institutionally power-equal field of study. {Management} has the most inequality, suggesting the affiliation of authors is highly correlated with their power in this field. To provide the reader with an intuition about power-inequality in author-citation networks, we show the induced subgraph~(i.e.,~subgraph constructed by picking a subset of nodes and edges between them) for the Management field where the nodes were partitioned by \textbf{(c)} gender and \textbf{(d)} prestige of their affiliation. In the gender-partitioned network, red (resp. blue) nodes represent female (resp. male) authors, and in the affiliation-partitioned network, they represent authors from top-ranked (resp. other) institutions. The induced subgraph is constructed from the ego-networks of a linked pair of red and blue nodes, which include all the nodes that cite or are cited by the linked pair of nodes, as well as the links between them. The links are assigned the same color as the as the node they point to (i.e.,~the author being cited). It can be clearly seen that the majority blue (male) group is cited disproportionately in the gender-partitioned network while the minority red group (affiliated with top-ranked institutes) is cited disproportionately more in the affiliation-partitioned network.
        \label{fig:power}
        } 
\end{figure}

We constructed author-citation networks from bibliometric data for six fields of study (see Methods).
An author-citation network at a discrete time $\timeValue = 1,2, \ldots$ is a directed graph $\graph_\timeValue = \{\nodeSet_\timeValue, \edgeSet_\timeValue\}$ where $\nodeSet_\timeValue$ is the set of authors and the $\edgeSet_\timeValue$ is the set of directed edges between them. An edge from author $u$ to $v$ exists if the latter cites any of the papers written by the former, i.e.,~the direction indicates the flow of information. Hence, the out-degree $\outDegree(v)$ of an author $v \in \nodeSet_\timeValue$ is the number of authors citing $v$, whereas the in-degree $\inDegree(v)$ is the number of people $v$ has cited up to and including time $\timeValue$. We use $\calBlue_\timeValue, \calRed_\timeValue \subset \nodeSet_\timeValue$ to represent a partition of the set of authors into two groups based on some attribute such as gender (female authors~$\calRed_\timeValue$ vs male authors~$\calBlue_\timeValue$) or prestige of their institutional affiliation (authors affiliated with top-ranked institutions~$\calRed_\timeValue$ vs the rest $\calBlue_\timeValue$). Further, the total in-degree (resp. out-degree) of red nodes is $\inDegree(\calRed_\timeValue) = \sum_{v \in{\calRed_\timeValue}}\inDegree(v)$  (resp. $\outDegree(\calRed_\timeValue) = \sum_{v \in{\calRed_\timeValue}}\outDegree(v)$), and the total in-degree of all nodes at time $\timeValue$ is $\degree_\timeValue = \sum_{v \in V_\timeValue} \inDegree(v)$.

Following \cite{avin2015homophily}, we use the average degree of a group as a proxy of its power in 
a graph. However, in directed graphs, we need to account for the in-degree $\inDegree(v)$ and the out-degree $\outDegree(v)$ for each author $v$.
We define the power of the red group~$\calRed_\timeValue$ 
at time $\timeValue$ as the ratio of the average out-degree to the average in-degree of red nodes at time $\timeValue$, and similarly for the blue group.  
In author-citation networks, the out-degree represents the amount of recognition an author receives from others, and the in-degree represents the attention the author pays to other authors. Thus, the power of a each group measures the average recognition the group receive relative to the average recognition the group gives to others. 
The power-inequality in the citation network at time $\timeValue$ quantifies the disparity in the normalized recognition the two groups receive as the ratio of the power of red group to the power of the blue group,
\begin{equation}
\label{eq:power_inequality}
\powerInequality_\timeValue 
= \frac{\outDegree(\calRed_\timeValue) }{\inDegree(\calRed_\timeValue)} \times \frac{\inDegree(\calBlue_\timeValue) }{\outDegree(\calBlue_\timeValue)}.
\end{equation}
When $\powerInequality_\timeValue <1$, the
red group has relatively less power  compared to the blue group at time $\timeValue$. On the other hand, when $\powerInequality_\timeValue > 1$, the red group has relatively more power than the blue group. Note that in undirected networks $\powerInequality_\timeValue = 1$ (at any time  $\timeValue$). Thus, our notion of power-inequality specifically captures the consequences of the directed edges in citation networks. 

Figure~\ref{fig:power} illustrates power-inequality. Fig.~\ref{fig:power}(a)-(b) show the induced subgraphs of author-citation networks in Management, partitioned by gender and affiliation respectively. Each subgraph represents the ego-networks of a linked pair of authors, one belonging to the red group and the other to the blue. A node's ego-network includes all nodes connected to it and all edges between them. Outgoing edges inherit the color of the author being cited. Although the size of the red group is the same in the two subgraphs, the different number of red and blue edges suggests disparities in citations: in Fig.~\ref{fig:power}(a) the majority (blue) group gets a disproportionately \textit{large}  number of citations (compared to its size) and has more power, whereas in Fig.~\ref{fig:power}(b), the minority (red) group does. We confirm that this observation is also valid at the full network scale by calculating power-inequality across several fields of study: Management, Psychology, Physics, Political Science, Economics, and Computer Science. Fig.~\ref{fig:power}(c)-(d) shows power-inequality over time in gender and affiliation networks. 
In gender-partitioned networks, women are the minority ($\calRed_{\timeValue}$ in Eq.~(\ref{eq:power_inequality})) and have less power than men ($\calBlue_{\timeValue}$). Political Science and Economics, despite some progress towards equality in recent years, have the greatest citations disparity. Psychology comes closest to gender parity in citations, and it is also most gender balanced, while in Political Science and Economics, women are a small minority with 34.20\% and 28.01\% of all authors.
In contrast, in affiliation-partitioned networks, authors from top-ranked institutions, who form the minority class, have much more power than the majority class~($\calBlue_\timeValue$). This is despite the fact that the size of the minority class in these networks is smaller than the minority class in gender networks (Table~\ref{tab:data_info}). This suggests that class imbalance is not the main cause of power-inequality.

\subsection*{Model of Network Growth}

To explore the origins of power-inequality, we present a model of a growing citation network, namely, the Directed Mixed Preferential Attachment~(DMPA) model. The DMPA model shows how disparities in citations can arise as a consequence of the interplay between the relative prevalence of the minority group, biases in individual preferences for citing similar or dissimilar authors  (homophily/heterophily), and the rate at which new authors join the citation network. 
The proposed model is powerful enough to produce a range of observed phenomena in citations networks, while also simple enough to be analytically tractable.

The model (see Methods for the details) 
describes the growth of a bi-populated  directed random graph consisting of two types of nodes. At every time step $\timeValue$, one of three types of edges is added. 
First, with probability $\probEventOne$, a new node (author) appears and an existing node cites it according to the following rules. The new node is assigned to red group with probability $\redBirthProb$~(otherwise, it is assigned to the blue group). A potential citing node is chosen from among the existing nodes by sampling proportional to their in-degrees plus a constant $\deltaIn$. The idea is that an author who is already citing many others is likely to cite the new author.
A new citation edge is created depending on the colors of nodes, governed by \textit{homophily}: if both nodes are red (resp.~blue), a link is created with probability $\homophilyRedEventOne$ (resp.~$\homophilyBlueEventOne$); otherwise, if the existing node is red (resp.~blue) and the new node is blue~(resp.~red), a link is added with probability $1-\homophilyRedEventOne$ ($1-\homophilyBlueEventOne$). 
Alternately, with probability $\probEventTwo$, a new node appears and cites an existing node, chosen from among existing nodes by sampling proportional to their out-degrees plus a constant $\deltaOut$. 
A new edge is created from the existing author to the new one based on homophily parameters $\homophilyBlueEventTwo, \homophilyRedEventTwo$, as above. 
Finally, with probability $1-\probEventOne- \probEventTwo$, a new citation edge appears between existing authors, resulting in \textit{densification} of the network. Both the citing and cited authors are chosen independently based on their in- and out-degree, and they are connected based on homophily parameters  $\homophilyRedEventThree, \homophilyBlueEventThree$.
For simplicity we assume that homophily for different edge creation events is the same and denoted by $\homophilyBlueAnyEvent, \homophilyRedAnyEvent$. If $\homophilyBlueAnyEvent > 0.5$, the blue group is homophilic,~i.e.,~blue nodes more likely to cite other blue nodes. On the other hand, if $\homophilyBlueAnyEvent < 0.5$, the blue group is heterophilic~i.e.,~blue nodes are more likely to cite red nodes. A similar interpretation holds for the red group (with the parameter~$\homophilyRedAnyEvent$).

The probability $\redBirthProb$ that a new node is assigned to the red group
determines class balance asymptotically: when $\redBirthProb < 0.5$, red is the minority; otherwise, it is the majority. We can independently set the minority and majority groups to be homophilic or heterophilic by adjusting  the parameters $\homophilyRedAnyEvent,\ \homophilyBlueAnyEvent$. In addition, we can control the growth dynamics of the graph by appropriately choosing $\probEventOne, \probEventTwo$, which determine the relative frequency with which new authors arrive and link to existing authors. For example, setting $1-\probEventOne-\probEventTwo > \probEventTwo > \probEventOne$ describes an author-citation network where most of the new citations form between existing authors, and less frequently when new authors cite (or are cited by) existing authors. Finally, $\deltaboth > 0$ controls the degree of preferential attachment: larger $\deltaboth$ corresponds to 
lower preference for linking to high-degree nodes.

These parameters enable our proposed DMPA model to asymptotically replicate structural properties of many real-world social networks, including
scale free degree distribution~\cite{barabasi1999emergence}, assortative mixing~\cite{mcpherson2001birds,newman2002assortative}, and, as we show here, power-inequalities.
The model generalizes previous models, specifically, a preferential attachment model for directed networks populated with nodes of a single type~\cite{bollobas2003directed} and a dynamic model for bi-populated undirected networks~\cite{avin2020mixed,avin2015homophily}. 
In addition to being limited to undirected networks, the latter model also fails to capture network densification via new edges between existing nodes. 

\subsection*{Analysis of the Model and Insights}
\subsubsection*{Theoretical Analysis} 
We analyze the proposed DMPA model to study the asymptotic values of power-inequality. First, we define some additional notation.
Let 
\begin{equation}
\label{eq:thetaTime}
\thetaIn_\timeValue = \frac{\inDegree(\calRed_\timeValue)}{\degree_\timeValue}, \quad \thetaOut_\timeValue = \frac{\outDegree(\calRed_\timeValue)}{\degree_\timeValue}.
\end{equation}
represent the fraction of the total in-degree and out-degree of the red group at time $\timeValue$.

\begin{theorem}[Almost sure convergence of DMPA model]
		\label{th:convergence_DMPA}
	Consider the DMPA model with the parameters $\redBirthProb, \probEventOne , \probEventTwo$, $\deltaIn = \deltaOut = \deltaboth$ and $\homophilyBlueAnyEvent^{(i)} = \homophilyBlueAnyEvent, \homophilyRedAnyEvent^{(i)} = \homophilyRedAnyEvent$ for $i  = 1,2,3$. Then, there exists $\deltaboth^{*} > 0$ such that, for all $\deltaboth > \deltaboth^{*}$, the state of the system $\thetaVec_\timeValue = [\thetaIn_\timeValue, \thetaOut_\timeValue]^T$
	converges to a unique value ${\thetaVec_{*} = [\thetaIn_{*}, \thetaOut_{*}]^T \in [0,1]^2}$  with probability $1$ as time $\timeValue \rightarrow \infty$.
\end{theorem}
Theorem~\ref{th:convergence_DMPA} states that the normalized sum of in-degrees~$\thetaIn_\timeValue$ and the normalized sum of out-degrees~$\thetaOut_\timeValue$  of red nodes converge to unique values (for sufficiently large $\deltaboth$) with probability~$1$. This allows us to use the unique asymptotic values to analyze the power-inequality. For simplicity, we assume that the homophily parameters do not differ across the three edge formation events and $\deltaIn = \deltaOut$, although these assumptions can be easily relaxed.  

The proof of Theorem~\ref{th:convergence_DMPA} is in the supplementary material. The key idea behind the proof is to first express the evolution $\thetaVec_\timeValue = [\thetaIn_\timeValue, \thetaOut_\timeValue]^T$ as a non-linear stochastic dynamical system. Then, by using stochastic averaging theory, we show that the non-linear stochastic dynamical system can be approximated by a deterministic ordinary differential equation~(ODE) of the form $\Dot{\thetaVec} = \nonLinFunction(\thetaVec) - \thetaVec$ where $\nonLinFunction$ is a contraction map~(for sufficiently large $\deltaboth$). Since $\nonLinFunction$ is a contraction map, it has a unique fixed point and, the sequence~$\thetaVec_\timeValue, \timeValue = 1, 2, \ldots$ converges to that fixed point (according to the Banach fixed point theorem). Since we know the contraction map $\nonLinFunction$ in closed form~(see supplementary material), the unique fixed point~$\thetaVec_{*} = [\thetaIn_{*}, \thetaOut_{*}]^T$ can be easily found via the recursion $\thetaVec_{k+1} = \nonLinFunction(\thetaVec_{k})$ with arbitrary initial condition $\thetaVec_{0} \in [0,1]^2$. Thus, Theorem~\ref{th:convergence_DMPA} implies that the power-inequality $\powerInequality_\timeValue$ in Eq.~\ref{eq:power_inequality} converges to 
$\powerInequality = {\thetaOut_{*}(1-\thetaIn_{*})}/{\thetaIn_{*}(1-\thetaOut_{*})}$. In addition, the asymptotic value of the power-inequality $\powerInequality$ for a specific parameter configuration can be found iteratively.
%
In contrast to \cite{avin2015homophily, avin2020mixed, bollobas2003directed}, which established convergence of $\thetaVec_\timeValue$ in expectation, we use tools from stochastic averaging theory to establish the almost sure convergence~(i.e.,~convergence with probability~$1$). 

\subsubsection*{Evaluation and Analysis of Power-inequality}

\begin{figure}
	\includegraphics[width=\columnwidth,  trim={0.15cm 0.3cm 0.2cm 0.1cm},clip]{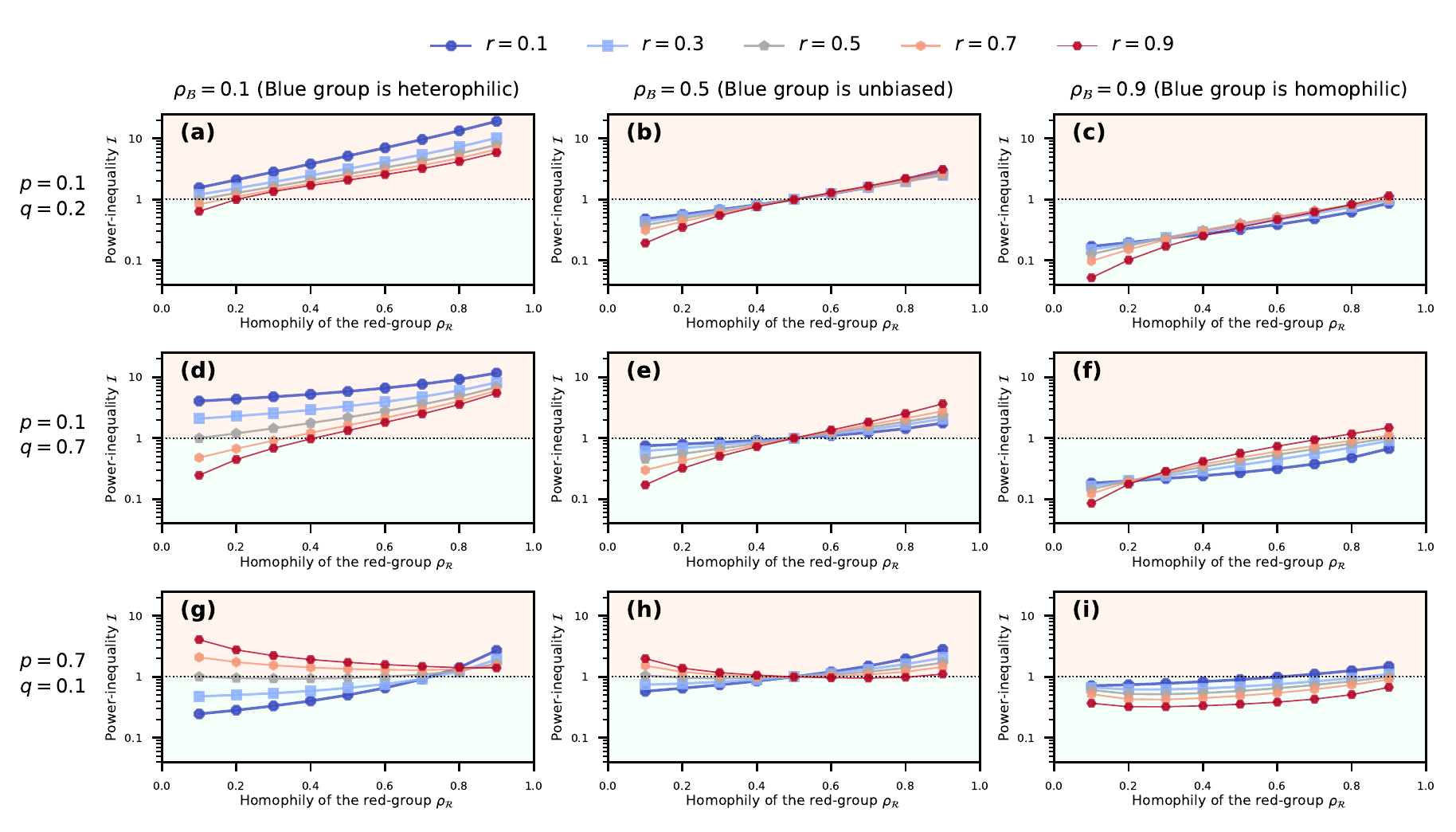}
	\caption{
	The figures display the power-inequality values for various parameter regimes of the proposed DMPA model.
	The three rows correspond to different values of the parameters $\probEventOne, \probEventTwo$ that capture the growth dynamics of the DMPA model. The three columns correspond to three different values of the homophily parameter of the blue group:~$\homophilyBlueAnyEvent = 0.1$~(heterophilic), $\homophilyBlueAnyEvent = 0.5$~(unbiased), $\homophilyBlueAnyEvent = 0.9$~(homophilic). In each subplot, lines in different colors correspond to various values of the parameter $\redBirthProb$ that determines the asymptotic class balance. In addition, $\deltaboth = 10$ for each case and another figure for $\deltaboth = 100$ is given in the supplementary material.
} 
	\label{fig:PowerInequalityTheoretical}
\end{figure}

Figure~\ref{fig:PowerInequalityTheoretical} displays the power-inequality evaluated for different parameter values using the iterative method discussed above. The results demonstrate that the DMPA model can produce a vast array of behaviors.  We now briefly present some of the insights from Fig.~\ref{fig:PowerInequalityTheoretical}.

Real-world networks typically densify as they grow~\cite{leskovec2007graph}, adding new edges between existing nodes. When a new node does appear, it connects to an existing node (i.e., citing an existing author), and it is rare for an existing author to cite a new one.  
The top row in Fig.~\ref{fig:PowerInequalityTheoretical} illustrates this scenario, with  $\probEventOne = 0.1,\ \probEventTwo = 0.2$. When the blue group is heterophilic~(Fig.~\ref{fig:PowerInequalityTheoretical}(a)),
i.e., it prefers to link to the red group, the red group is more powerful
whenever it is considerably more homophilic than the blue group. 
Under these conditions, power-inequality $\powerInequality$ increases exponentially with red group's homophily~$\homophilyRedAnyEvent$. Surprisingly, $\powerInequality$ decreases 
with $\redBirthProb$, meaning that the smaller the red group, the more powerful it is. This unusual behavior is especially apparent when the red group is highly homophilic~($\homophilyRedAnyEvent = 0.9$). Therefore, when one group is highly homophilic and the other highly heterophilic, the heterophilic group faces a disparity in power, and increasing its size does not ameliorate the disparity. For an intuition, consider membership in an exclusive club: those who are already in this selective group prefer to associate with others in it (they are homophilic), and those who are not also prefer to associate with the exclusive group (they are heterophilic); therefore, making the exclusive group smaller serves to make it even more powerful. 
As a result, the most effective approach for ameliorating power-inequality is to alter individual preferences by changing the homophily parameters~$\homophilyBlueAnyEvent, \homophilyRedAnyEvent$ rather than sizes of groups. 
On the other hand, when the blue group is unbiased (Fig.~\ref{fig:PowerInequalityTheoretical}(b)),
the red group is more powerful 
when $\homophilyRedAnyEvent>\homophilyBlueAnyEvent$, 
and its power increases with group size.  There is less variation in power-inequality (compared to the top-left panel), suggesting that power disparity is driven by the difference of the homophily parameters of the two groups. Finally, when the blue group is highly homophilic (Fig.~\ref{fig:PowerInequalityTheoretical}(c))
it is more powerful than the red group~(i.e.,~$\powerInequality \leq 1$), 
and power-inequality decreases with~$\homophilyRedAnyEvent$.

The middle row in Fig.~\ref{fig:PowerInequalityTheoretical} corresponds to $\probEventOne < 1-\probEventOne-\probEventTwo < \probEventTwo$, which describes fast growing phase of the citations network, e.g., during the establishment of a new research field, where an influx of new authors creates many new citations to existing authors. Most trends observed in the first setting~($\probEventOne = 0.1, \probEventTwo = 0.2$) still hold; however, the effect of group size, determined by $\redBirthProb$, is more pronounced, since more new nodes are added to the network. 

Finally, the bottom row corresponds to the case $\probEventOne > 1-\probEventOne-\probEventTwo > \probEventTwo$, which models the case where the citations network grows primarily via existing authors citing new authors rather than other existing authors. 

The worst case values of power-inequality (corresponding to $\homophilyBlueAnyEvent = 0.1, \homophilyRedAnyEvent = 0.9$ and $\homophilyBlueAnyEvent = 0.9, \homophilyRedAnyEvent = 0.1$) become less severe when moving from top row to bottom row; this suggests that frequently bringing in new authors is one potential way to battle the disparities of power-inequality in the presence of extreme homophilic biases. 


So far, our numerical results indicate how parameters of the DMPA model affect power-inequality for a fixed value of $\deltaboth = 10$. 
To reduce the effects of preferential attachment, we re-evaluated power-inequality with $\deltaboth = 100$ (see Supplementary Fig.~\ref{fig:PowerInequalityTheoretical_delta100}). The results show that preferential attachment amplifies power-inequality, especially when there is a high degree of class imbalance i.e.,~the effect of $\deltaboth$ on power-inequality $\powerInequality$ is larger when $\redBirthProb \ll 0.5$ or $\redBirthProb \gg 0.5$. Thus, controlling $\deltaboth$ is a useful strategy to ameliorate power-inequality in the presence of extreme class imbalance.

\begin{figure}
    \centering
    \includegraphics[width=1\columnwidth]{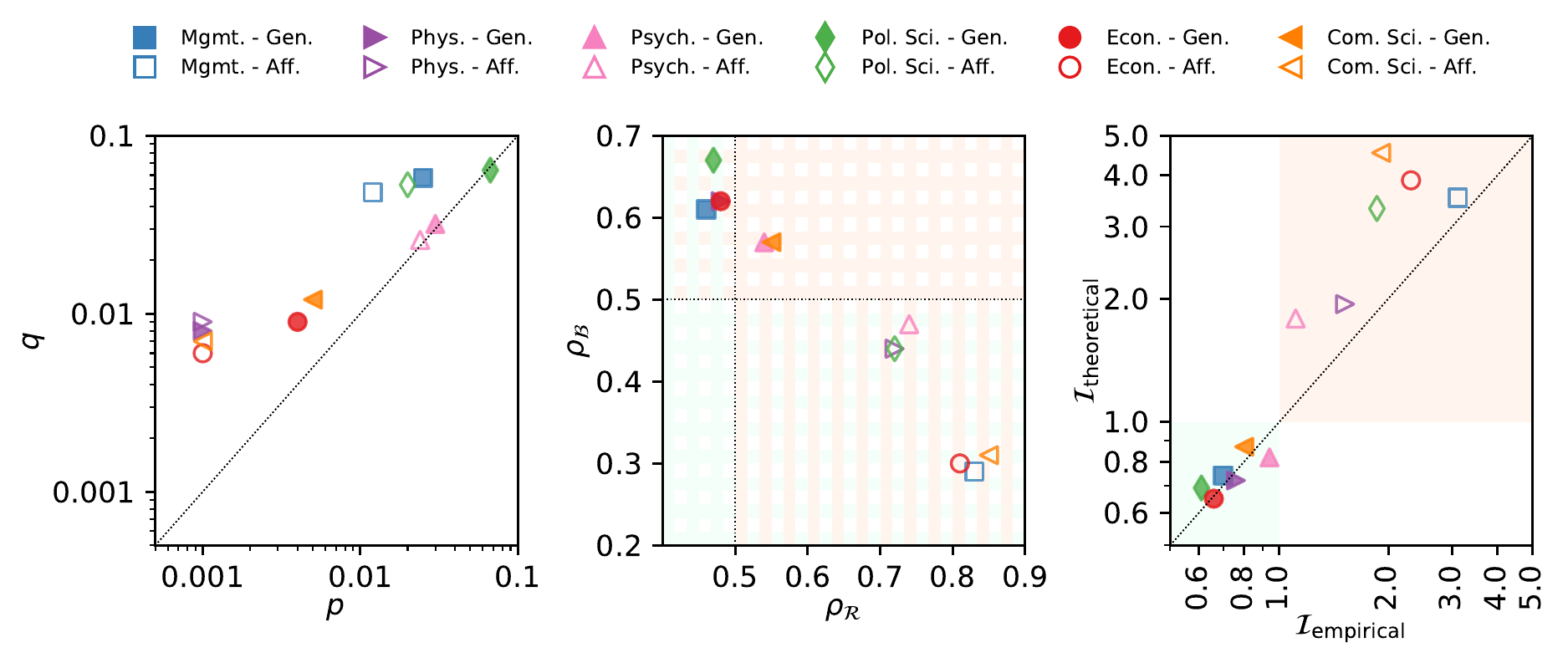}
    \caption{Empirically estimated parameters of the DMPA model using the gender-partitioned networks (filled markers) and affiliation-partitioned networks (open markers). The exact values are listed in Table~\ref{tab:params}. The leftmost plot suggests that the probability of a new node citing an existing node is larger than the probability of an existing node citing a new node (i.e.,~$\probEventTwo > \probEventOne$) for most fields of studies. The middle plot shows that minority female authors (red group) are heterophilic ($\homophilyRedAnyEvent < 0.5$) while the majority male authors are homophilic ($\homophilyBlueAnyEvent > 0.5$). The opposite is observed in affiliation networks; the authors affiliated with top-ranked universities~(i.e.,~minority) are homophilic while the others are heterophilic.  The rightmost plot shows that empirically estimated values of the power-inequality are in close agreement with the values obtained using the DMPA model. This shows that the DMPA model can represent how the power-inequality emerge in real-world networks. Moreover, combining the empirically estimated parameters of the DMPA model with its theoretical analysis provides insight on strategies that can be used to mitigate the power-inequality.}
    \label{fig:estimated_parameters}
\end{figure}

\subsubsection*{Connecting to the Real-World Networks}
To relate the key insights from the theoretical analysis of the DMPA model to real-world author-citation networks, we estimated the parameters of the model (see Methods) from data. The empirically estimated parameters and values of power-inequality calculated with those parameters are shown in Fig.~\ref{fig:estimated_parameters} (full set of parameters are listed in Table~\ref{tab:params}).
In gender-partitioned citation networks (filled symbols in Fig.~\ref{fig:estimated_parameters}), the theoretically calculated values of power-inequality are in close agreement to their empirical values that are measured from citation networks using Eq.~\ref{eq:power_inequality}. 
Both the empirical and theoretically calculated values of power-inequality confirm that the minority group (female authors) experience a dispartiy in recognition. 
 Note that $\probEventOne < \probEventTwo \ll 1-\probEventOne-\probEventTwo$ in all cases except Political Science where $\probEventTwo < \probEventOne \ll 1-\probEventOne-\probEventTwo$. Therefore, densification of the citation network through edge formation between existing authors is the most frequent event, and new authors join less frequently. Moreover, the blue group (representing male authors) is homophilic ($\homophilyBlueAnyEvent > 0.5$ in  Fig.~\ref{fig:estimated_parameters}). Hence, gender-partitioned citation networks generally fall into the scenario captured by the top-right panel of Fig.~\ref{fig:PowerInequalityTheoretical}. 

In affiliation-partitioned citation networks, the red group (authors from top-ranked institutions) is in the minority ($\redBirthProb < 0.5$) in each field. As shown in Fig.~\ref{fig:estimated_parameters}, the minority has more power according to both the theoretical and empirical values of power-inequality. Although theoretically calculated values overestimate power-inequality, they have the same ranking as the empirical values in all fields except Computer Science and Management. The minority group is strongly homophilic~($\homophilyRedAnyEvent > 0.7$ in all cases) and the majority is heterophilic~($\homophilyBlueAnyEvent < 0.5$ in all cases). Further, note that $\probEventOne < \probEventTwo \ll 1-\probEventOne-\probEventTwo$; hence, the affiliation-partitioned networks correspond to the scenario shown in the top-left panel of Fig.~\ref{fig:PowerInequalityTheoretical}. 

The estimated preferential attachment parameter $\deltaboth$ is $1000$ for most networks except for {Psychology}, {Economics}, and {Computer Science} in gender-partitioned networks and {Political Science} in affiliation-partitioned networks. The difference between homophily $|\homophilyBlueAnyEvent - \homophilyRedAnyEvent|$ is smaller for these networks than the others, suggesting that preferential attachment plays a more important rule than homophily in shaping power-inequality in these networks. 

According to our empirical results, {psychology} is the most egalitarian research field, with power-inequality values closest to one, and some of the smallest gaps between homophily parameters for the minority and the majority classes. In addition, half of the authors in the gender-partitioned network belong to the red group, as well as quarter of the authors in the affiliation-partitioned network, a larger fraction compared to other fields.
In contrast, {Political Science} and Management have the highest empirical power-inequality for gender and affiliation networks respectively. 
The poor agreement between theoretical and empirical values of power-inequality for Computer Science could be due to the strong temporal variability of the citation networks in Computer Science. 
There appears to be a weak interaction between gender and affiliation. Affiliation with a top-ranked institutions is weakly correlated with being male~(Pearson correlation ranges from  $r=0.01$ to $r=0.04$, $p<0.001$) in all fields of study except for Psychology, where female authors are slightly more likely to be affiliated with high-ranked institutions (Pearson coefficient $r=-0.01$, $p<0.001$), and Management, where correlation is not statistically significant. 


\begin{figure*}[h!] 
        \centering
        \begin{subfigure}{0.45\textwidth}
            \centering
            \includegraphics[width=\textwidth]{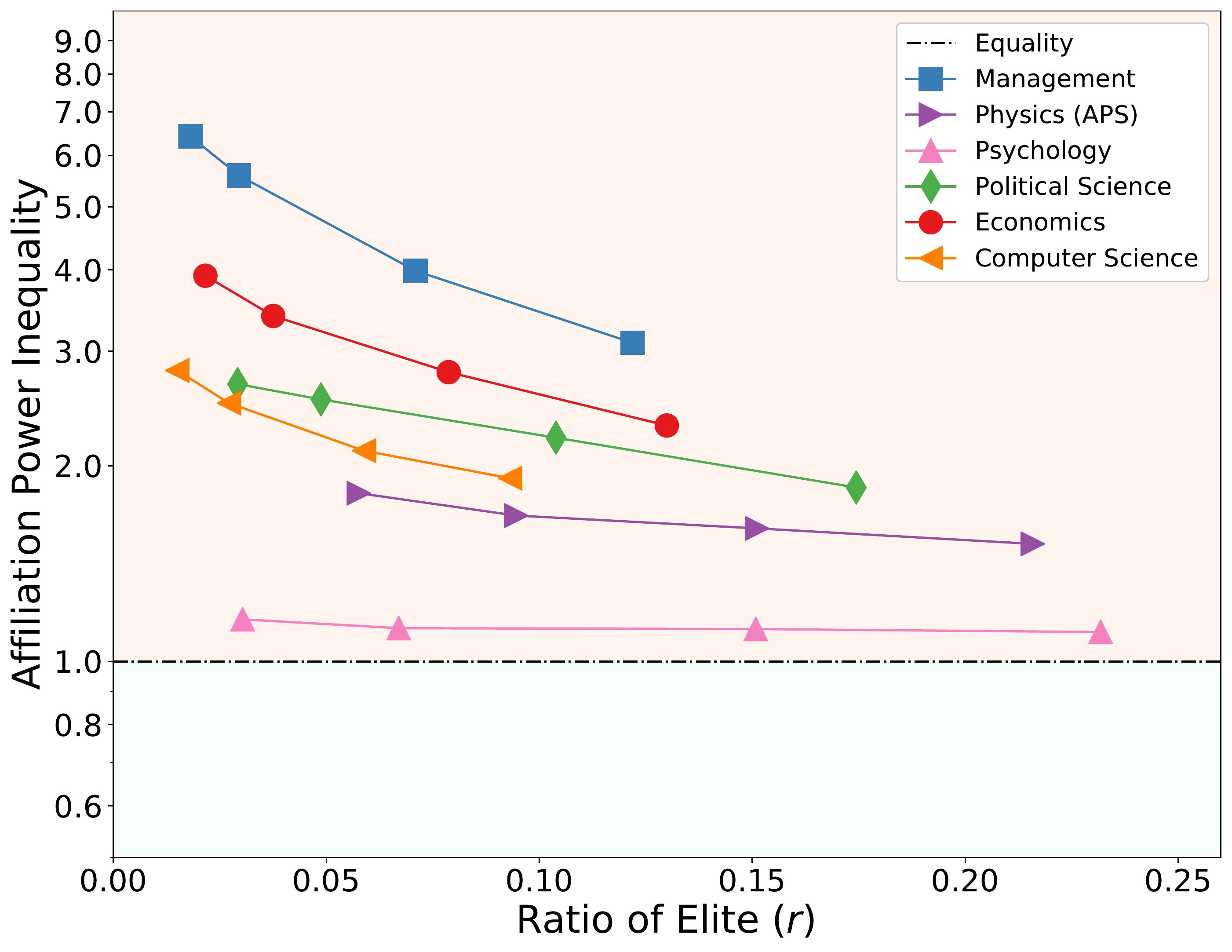}
            \caption{Affiliation Power-inequality}
        \end{subfigure} \hspace{0.5cm}
        \begin{subfigure}{0.45\textwidth}
            \includegraphics[width=\textwidth]{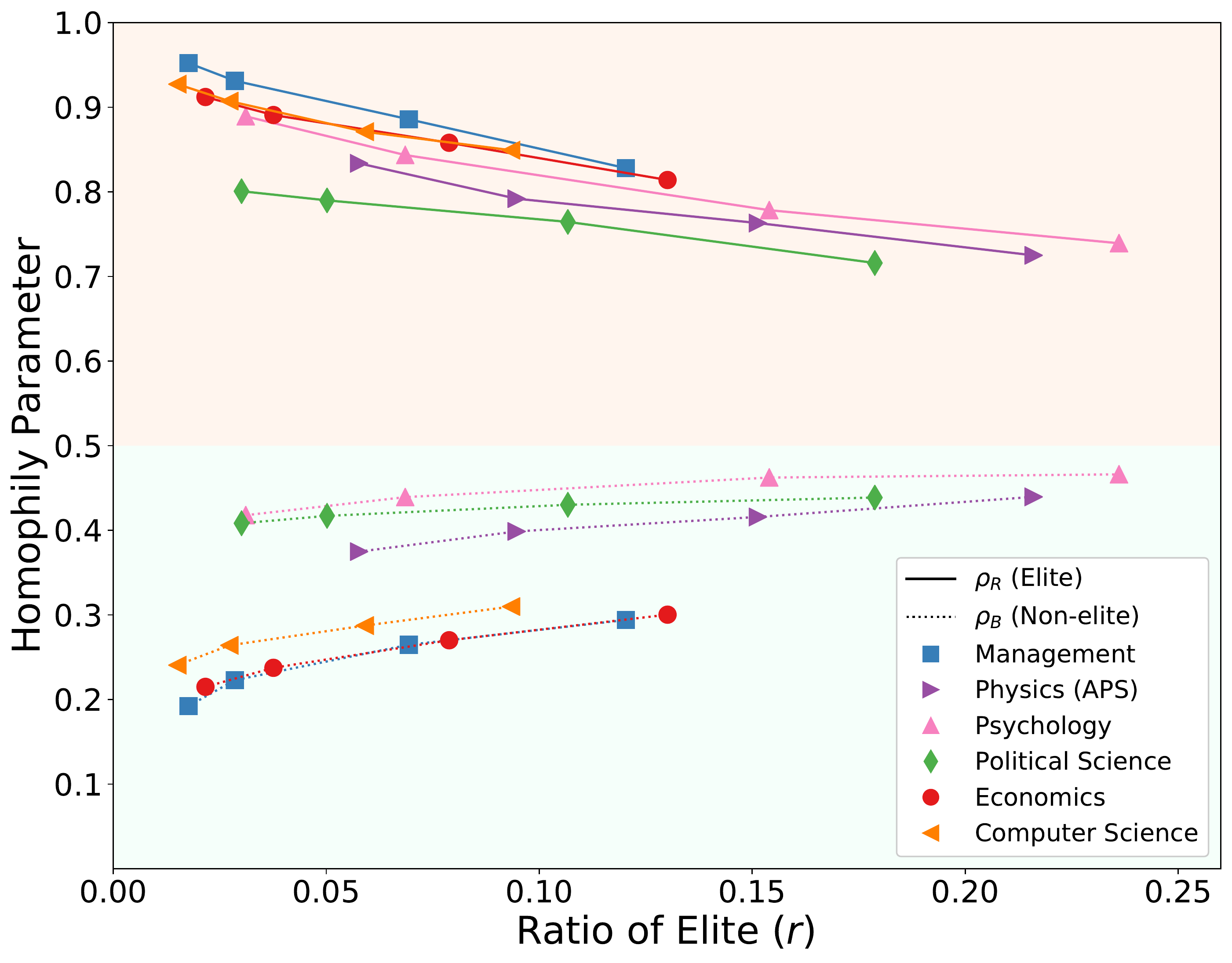}
            \caption{Homophily parameters}
        \end{subfigure}
    \caption{The figures illustrate how (a)~power inequality and, (b)~homophily parameters of the two groups, vary with the minority group size in affiliation networks by defining ``top-ranked universities" to be the top 10,20,50 and 100 universities in \emph{Shanghai University Rankings(SUR).}
    We use $\deltaboth$ from Table \ref{tab:params} for each field of study. The figure shows that smaller the minority group is, the more powerful it is, as a consequence of minority and majority groups being increasingly homophilic and heterophilic, respectively. This observation agrees with the predictions of the DMPA model shown in Fig.~\ref{fig:PowerInequalityTheoretical}(a)
    \label{fig:elite_range}}
\end{figure*} 

\subsection*{Smaller Elite Group Has More Power}
To assess the generalizability of our approach, we estimated power-inequality of the most prestigious institutions by changing the definition of ``top-ranked'' from the top-100 institutions to the top-50, top-20, and top-10 institutions {listed in \textit{Shanghai University Rankings} (see Supplementary Information for details)}. As the number of institutions considered as ``top-ranked'' shrinks, so does the minority fraction of authors belonging to those institutions ($\redBirthProb$). Figure~\ref{fig:elite_range}(a) reports power-inequality as a function of the minority fraction $\redBirthProb$ for these different cases. For Management, Economics, and Computer Science, the relative power of the minority grows significantly as its size shrinks; when authors from the top-10 institutions are considered instead of the top-100 institutions, the power-inequality almost doubles. Fields such as Psychology, and to some extent Physics, are more egalitarian and do not show a significant increase in the power of the minority as its size shrinks. The estimated homophily parameters (Fig.~\ref{fig:elite_range}(b)) suggest that the smaller minorities are more homophilic and prefer to cite themselves more, whereas the majority group is more strongly heterophilic.

\subsection*{Mitigating Power-inequality in Author-Citation Networks} 

We now discuss insights the DMPA model offers for reducing the disparity in citations.
Recall that 
Fig.~\ref{fig:PowerInequalityTheoretical}(c) corresponds to gender-partitioned citation networks. Analysis suggests that simply increasing the size of the minority group is not sufficient to reduce the disparity, as power-inequality exists even when group sizes are equal~($\redBirthProb = 0.5$). Reducing the homophily of each group, so that the groups show no preference for citing their own members, is a more effective strategy: it reduces the disparity of citations even with unequal group sizes, as seen from the middle column of Fig.~\ref{fig:PowerInequalityTheoretical}. This  explains why Psychology and Computer Science have power-inequality values closer to 1---both groups in those fields have more similar homophily parameters as seen in Table~\ref{tab:params}. On the other hand, Political Science, Economics, Management, and Physics have the most disparity in citations, and in all four cases, the minority red group is heterophilic and the majority blue group is homophilic. Additionally, recall that the worst-case~(i.e.,~when difference between $\homophilyBlueAnyEvent$ and $\homophilyRedAnyEvent$ is close to 1) power-inequality is largest in the top row and smallest in the bottom row of  Fig.~\ref{fig:PowerInequalityTheoretical}. Therefore, encouraging new authors to join a field of study (i.e.,~making $\probEventOne, \probEventTwo \gg 1-\probEventOne - \probEventTwo$) and especially, providing them the opportunity to have an audience among established authors~(i.e.,~$\probEventOne > \probEventTwo \gg 1-\probEventOne - \probEventTwo$) also reduces the  power-inequality. Finally, in cases where neither homophily, growth dynamics~($\probEventOne, \probEventTwo$), nor group size ($\redBirthProb$) can be controlled, power-inequality could be reduced by decreasing the importance of preferential attachment (i.e., increasing $\deltaboth$). 
In practice, this could be achieved by providing support to researchers and authors in a manner that is not solely dependent on their past performance. 

In affiliation networks, which correspond to 
Fig.~\ref{fig:PowerInequalityTheoretical}(a), 
power-inequality is a clear consequence of homophily: the majority is heterophilic ($\homophilyRedAnyEvent<0.5$) and the minority is strongly homophilic ($\homophilyRedAnyEvent > 0.5$). Interestingly, Fig.~\ref{fig:PowerInequalityTheoretical}(a)
suggests that the smaller the minority, 
the larger the power-inequality.
This observation is confirmed by Fig.~\ref{fig:elite_range}, which shows power-inequality for different minority group sizes (corresponding to different values of $\redBirthProb$). 
This suggests that the best way to alleviate citation disparities with respect to institutional prestige is to make the citations free of the influence of author affiliation, for example, by simply highlighting the names of authors and not their affiliations in the title page of scientific publications. In addition, as with gender networks, encouraging new authors to join a research field and providing them an audience among established authors~(i.e.,~$\probEventOne > \probEventTwo \gg 1-\probEventOne - \probEventTwo$), as well as reducing preferential attachment, also mitigates power-inequality.

\section*{Methods}
\subsection*{Data}
\definecolor{Grey}{gray}{0.9}
\begin{table}[tb]
    \centering
    \begin{tabular}{|c||c|c|c||c|c|c|}
        \hline 
        \multirow{2}{*}{\textit{Field of study}} &        
        \multicolumn{3}{c||}{\cellcolor{Grey}  \textit{Gender-partitioned Network}} & \multicolumn{3}{c|}{\cellcolor{Grey}  \textit{Affiliation-partitioned Network}} \\
         \cline{2-7}
            & \# Nodes & Density & \% Female & \# Nodes & Density & \% Elite \\
         \hline 
         Management & $111K$ & {9.9$\times 10^{-5}$} & $35.19$ & $53K$ & {3.0$\times 10^{-4}$} & $12.19$ \\ 
         \hline 
         Physics & $164K$ & {6.4$\times 10^{-4}$}& $16.39$ & $156K$ & {6.3$\times 10^{-4}$} & $21.59$ \\
         \hline 
         Psychology & $873K$ & {1.7$\times 10^{-5}$} & $49.65$ & $667K$ & {2.9$\times 10^{-5}$} & $23.18$ \\ 
         \hline 
         Political Science & $1.03M$ & {7.0$\times 10^{-6}$} & $34.20$ & $204K$ & {6.4$\times 10^{-5}$} & $17.44$ \\ 
         \hline 
         Economics & $1.3M$ & {5.6$\times 10^{-5}$} & $28.01$ & $723K$ & {1.8$\times 10^{-4}$} & $13.00$ \\ 
         \hline 
         Computer Science & $7.62M$ & {7.5$\times 10^{-6}$} & $25.79$ & $3.27M$ & {3.9$\times 10^{-5}$} & $9.31$\\ 
         \hline 
    \end{tabular}
    \vspace{0.1in}
    \caption{Information about the data. Data for all fields of study, except Physics, came from Microsoft Academic Graph, and Physics data was provided by the American Physical Society. The number of authors with known gender is larger than number of authors with known affiliation. The affiliation network has higher density, potentially confounded by the fact that authors with known affiliation are more active in publishing and citing other authors. 
    }
    \label{tab:data_info}
\end{table}

We used the Microsoft Academic Graph (MAG) API to collect metadata from papers published in Computer Science, Management, Psychology, Political Science, and Economics since 1990. We also collected metadata of papers published in the journals of the American Physical Society (APS). We consider the authors of these papers to represent the field of Physics.
To study gender disparities, we used the \textit{Gender API} ({\url{https://gender-api.com/}) to get an author's gender from their name. We eliminated authors whose first names are not recognized by the \textit{Gender API}. 
For each year, we constructed a directed author-citation network where an edge from $u$ to $v$ with weight $w$ indicates author $u$ cited author $v$ $w$ times in the papers $u$ published that year. 
The majority class are male authors, and the remaining authors  form the minority class (female authors).

We also extracted the author's most recent affiliation, removing authors whose affiliations were unavailable. We used \textit{Shanghai University Rankings} (SUR, \url{https://www.kaggle.com/mylesoneill/world-university-rankings})  to identify authors affiliated with prestigious institutions that were among the top-100 institutions. We constructed author citation networks the same way as above, but now, the authors affiliated with prestigious (top-100) institutions form the minority class, and authors affiliated with the remaining  institutions are the majority class. 
Table \ref{tab:data_info} presents statistics of these citation networks.  

\subsection*{Model}
We propose a model of growing bi-populated directed networks that captures the key elements of real-world dynamics, while being simple enough for theoretical analysis.
The proposed model has parameters defining class balance ($\redBirthProb$), growth dynamics of the network ($\probEventOne,\ \probEventTwo$), homophily ($\matrixEventOne,\ \matrixEventTwo , \matrixEventThree$) and preferential attachment ($\deltaIn,\ \deltaOut$).

The model parameters $\matrixAnyEvent _i, i \in \{1,2,3\}$ are  matrices of the form
\begin{equation}
\matrixAnyEvent_{i} = \begin{bmatrix}
\homophilyBlueAnyEvent^{(i)} & 1 - \homophilyRedAnyEvent^{(i)} \\
1 - \homophilyBlueAnyEvent^{(i)} & \homophilyRedAnyEvent^{(i)} 
\end{bmatrix}
\end{equation} 

The network at time $\timeValue$ is denoted by $\graph_\timeValue = (\nodeSet_\timeValue, \edgeSet_\timeValue)$ where the set of nodes $\nodeSet_\timeValue$ can be partitioned into a set of blue nodes~$\calBlue_\timeValue$ and red nodes~$\calRed_\timeValue$. 
The initial graph $\graph_0$ corresponds to $2\times 2$ adjacency matrix containing all 1's, though $\graph_0$ could be any arbitrary matrix and the asymptotic state of the model does not depend on it.

\begin{figure}
    \centering
    \includegraphics[width=0.65\linewidth]{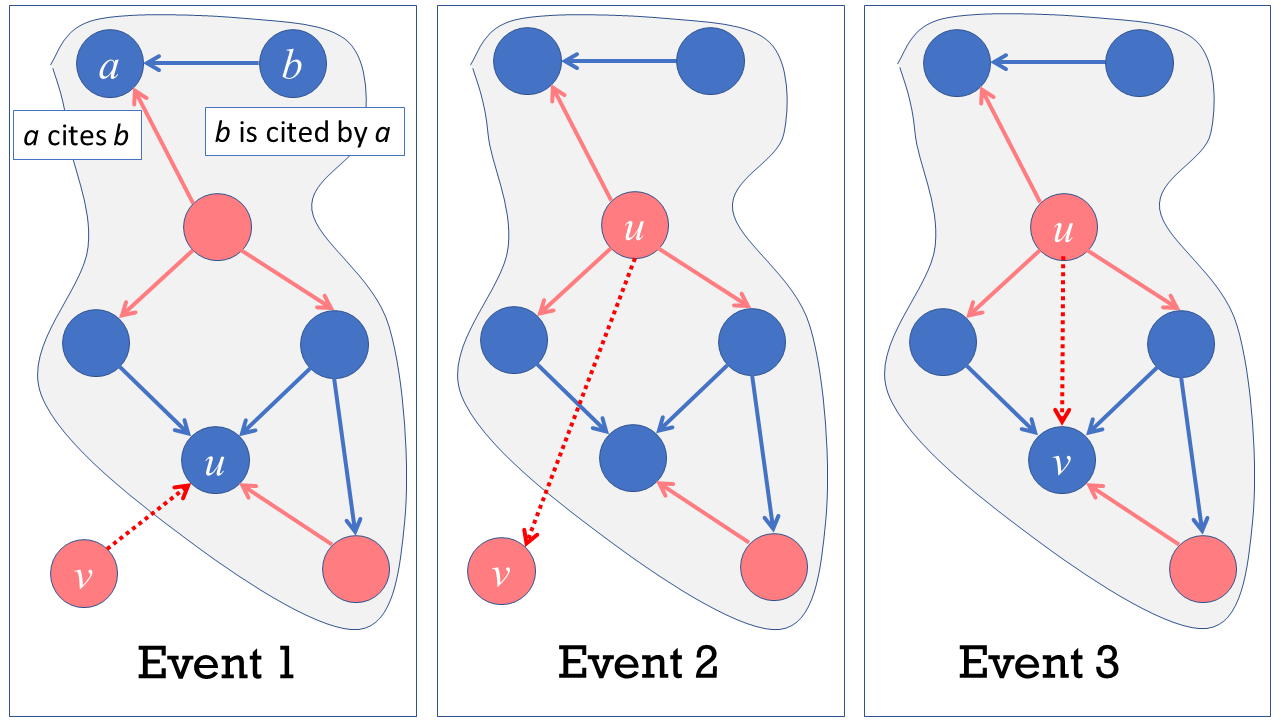}
    \caption{Citation edge creation events considered by the DMPA model of the growth of bi-populated directed networks. The first two events correspond to the appearance of a new node. A new directed edge is created when an existing node cites a new node (Event 1) or the new node cites an existing node (Event 2). Event 3 shows densification of the network via a new edge appearing between existing nodes. 
    }
    \label{fig:model}
\end{figure}

At each time step $\timeValue$, one of three events happens, as shown in Fig.~\ref{fig:model}:
\begin{enumerate}[itemsep=0.1em, parsep = 0.2em]
\item Event 1 (with probability $\probEventOne$): An existing node follows the new node
\begin{enumerate}[label=\roman*., itemsep=0.05em, topsep=0pt, parsep = 0.2em]
    \item {\bf Minority - majority partition:} A new node $v_\timeValue$  appears and is assigned color red with probability $\redBirthProb$ and blue with probability $1-\redBirthProb$.
	
	\item {\bf Preferential attachment:} 
	An existing node $u_\timeValue \in \nodeSet_\timeValue$ is chosen by sampling with probability $\propto \inDegree(u_\timeValue) + \deltaIn$. 
	
	\item {\bf Homophily:} If both $u_\timeValue, v_\timeValue$ are red (resp. blue), an edge $(v_\timeValue, u_\timeValue)$ is added with probability $\homophilyRedEventOne$ (resp. $\homophilyBlueEventOne$). Otherwise, if $u_\timeValue$ is red (resp. blue) and $v_\timeValue$ is blue (resp. red), a link $(v_\timeValue, u_\timeValue)$ is added with probability $1 - \homophilyRedEventOne$ (resp. $1 - \homophilyBlueEventOne$). 
	
	\item Above steps ii, iii are repeated until an outgoing edge is added to $v_\timeValue$. 
\end{enumerate}

\item Event 2 (with probability $\probEventTwo$): The new node follows an existing node
\begin{enumerate}[label=\roman*., itemsep=0.05em, topsep=0pt, parsep = 0.2em]
	\item {\bf Minority - majority partition:} as above. 
	
	\item {\bf Preferential attachment:} An existing node $u_\timeValue \in \nodeSet_\timeValue$ is chosen by sampling with probability  $\propto \outDegree(u_\timeValue) + \deltaOut$. 
	
	\item {\bf Homophily:} as above. 
	
	\item Above steps ii, iii are repeated until an incoming edge is added to $v_\timeValue$.
\end{enumerate}

\item Event 3 (with probability $1 - \probEventOne  - \probEventTwo$): Densification through new edges between existing nodes
\begin{enumerate}[label=\roman*., itemsep=0.05em, topsep=0pt, parsep = 0.2em]
	\item {\bf Preferential attachment:} An existing node $u_\timeValue \in \nodeSet_\timeValue$ is chosen by sampling with probability  $\propto \outDegree(u_\timeValue) + \deltaOut$ and another node $v_\timeValue \in \nodeSet_\timeValue$ is chosen by sampling with probability  $\propto \inDegree(v_\timeValue) + \deltaIn$. 
	
	\item {\bf Homophily:} as above. 
	
	\item Above steps i, ii are repeated until a new edge is added to the graph.
\end{enumerate}
\end{enumerate}

\subsubsection*{Estimating model parameters from data}
The bibliometric data usually includes only the year of publication, and as a result, we do not know the order in which citations edges appear at a finer temporal resolution. To better match the conditions of the DMPA model, we create an ordered list of edges $E_{ord}$ as follows. We initialize a set of nodes $S$ with one of the nodes from the largest connected component of the citation graph for 1990. Then in each iteration of the algorithm (for each year), we shuffle all the edges that have at least one end in set $S$, and append all of them to the end of list $E_{ord}$. Then we update set $S$ with the nodes covered by this newly added batch of edges. We continue this process until there are no more edges from that year that are connected to $S$. We then drop all the remaining edges from that year that are not part of the connected component and continue the procedure for the following year. 

We use the nodes and edges in the ordered edge list to estimate all parameters of the DMPA model directly from the data, except for the preferential attachment parameter $\deltaboth$. We estimate the preferential attachment parameter through hyper-parameter tuning. Specifically, we use several different values of $\deltaboth$ to estimate the remaining parameters of the DMPA model, and use them to calculate the theoretical value of power-inequality. We then choose the value of $\deltaboth$ that leads to power-inequality closest to its empirical value. The parameter estimation procedure is not sensitive to the choice of the initial node to populate the seed set, nor the order in which edges are added. Running the procedure multiple times yields parameter estimates with standard deviation less than $10^{-4}$. Details of parameter estimation are described in the Supplementary Information.

\section*{Conclusion}


We studied inequalities in scientific citations that lead one group of authors of scientific publications---women or researchers from less prestigious institutions---to receive less recognition for their work than the advantaged group (men, researchers from the top-ranked institutions). 
%
To explain these disparities, we proposed a novel model (Directed Mixed Preferential Attachment model)  of the growth of citations networks which captures biases in authors' individual preferences to cite others who are similar to them, well-recognized or highly active. The model also specifies the relative frequency with which new authors join. Its predictions align closely with empirical observations, indicating that the model's mechanism is useful for understanding gender and prestige-based disparities in citations.

Are these disparities in recognition of authors detrimental to science? Some have argued that inequalities are ingrained in the  structure of scientific rewards, which are skewed to channel the biggest rewards to a few top performers through mechanisms such as cumulative advantage~\cite{xie2014undemocracy}. These skewed reward mechanisms benefit science by incentivizing researchers to produce outstanding work, which earns them placement at the more prestigious institutions.
However, recent literature has called the link between prestige and merit into question. Research has shown that inequalities in individual productivity arise due to the cumulative benefits of early career placement rather than individual merit~\cite{way2019productivity}. Moreover, prestige, rather than the quality of scientific ideas, affects how quickly and widely they spread~\cite{morgan2018prestige}. 
Structural inequalities in recognition also likely contribute to the gender gap in science. 
Although women researchers publish at a similar rate as men, they tend to leave academia sooner~\cite{huang2020historical}, 
potentially because citations disparities lower their scientific impact.
Since hiring and promotion  decisions depend on the metrics of impact, the power-inequality we demonstrate could fundamentally limit women's opportunities for professional advancement regardless of their inherent merit, motivation, or ability. 
Structural inequalities, therefore, reduce the pool of available talent and decrease the diversity of the scientific workforce, which limits innovation by reducing the creativity and productivity in research~\cite{page2007power, smith2017diversity, Woolley2010}. 
To mitigate the effects of power-inequality in citations, our analysis suggests that 
reducing biases in citation patterns, e.g., by 
encouraging authors to cite researchers from underrepresented groups, is more effective than just increasing the size of the underrepresented group alone, e.g., by hiring. In addition, bringing in new authors and allowing them to have an accessible platform among established authors, and also reducing preferential attachment by providing some degree of support for new authors, are some other helpful strategies for reducing inequality. 

Finally, we also note that citations disparities may be amplified algorithmically by academic search engines, such as Google Scholar, which highlight authors and papers with most citations. To reduce bias, search engines should diversify search results or highlight papers by non-privileged groups. Academic publishers can also play an important role in reducing power-inequality, for example, by offering incentives to diversify citations and highlight new authors. Our work provides insights into measures and metrics publishers can use to reduce structural inequalities. 



\appendix

\section{Supplementary Information}
\subsection{Details on the Datasets}
\label{sec:data_lim} 
We use bibliographic data from Microsoft Academic Graph (MAG) and American Physiological Society (APS). The APS data was used to analyze the citations in the field of
Physics, and MAG was used to study the remaining fields, namely, Management, Psychology, Political Science, Economics and Computer Science. In this section, we discuss potential biases introduced by our data collection and processing methods as well as their potential impact on our analysis. 

To extract an author's gender, we use the Gender-API (\url{https://www.gender-api.com/}). This state-of-the-art API infers the gender based on the author's name. However, it fails to recognize the gender of authors who use initials instead of their full first names (e.g. ``J. Doe'' instead of ``John Doe'') and some Chinese and Korean names. 
Computer Science has the lowest coverage for gender~(i.e.,~the fraction of authors whose gender was inferred with Gender-API), while the coverage improves over time and reaches $80\%$. For other fields of study, the gender of at least $85\%$ of authors was known. We removed all authors whose gender was unknown.

MAG extracts author affiliations from publications and links it to a unique ID. Due to the challenges of automatically extracting affiliations from publications, the coverage for this field in the MAG data is low.  In contrast, affiliations in the APS data are specified by authors and stored as strings. We use string processing and normalization to map the affiliation to a unique name, which is then mapped to the ranking. Specifically, taking the full address of author's institutional affiliation, we remove stop words (e.g., ``the'') and convert all characters to lower-case. After separating the country, city, institution and department names of the affiliation (using python packages such as geonamescache), we string matched words \textit{``university''}, \textit{``institute''}, \textit{``laboratory''} or \textit{``college''} in different languages (e.g., \textit{``università''}, \textit{``universidad''}, \textit{``institut''}). We then matched the extracted and normalized institution names to those listed in Shanghai University Ranking. The coverage of affiliation for the APS data (Physics) is more than $97\%$. 
We consider authors without affiliation as not being affiliated with top-ranked institutions. 
While this could potentially bias the data (for example, when top-ranked authors' affiliation is not known), Fig.~\ref{fig:elite_range} provides evidence that this is not the case. The monotonicity of the trend for power-inequality suggests that the affiliations data is not systematically biased.

\begin{figure}[h!]
    \centering
    \includegraphics[width=0.7\columnwidth]{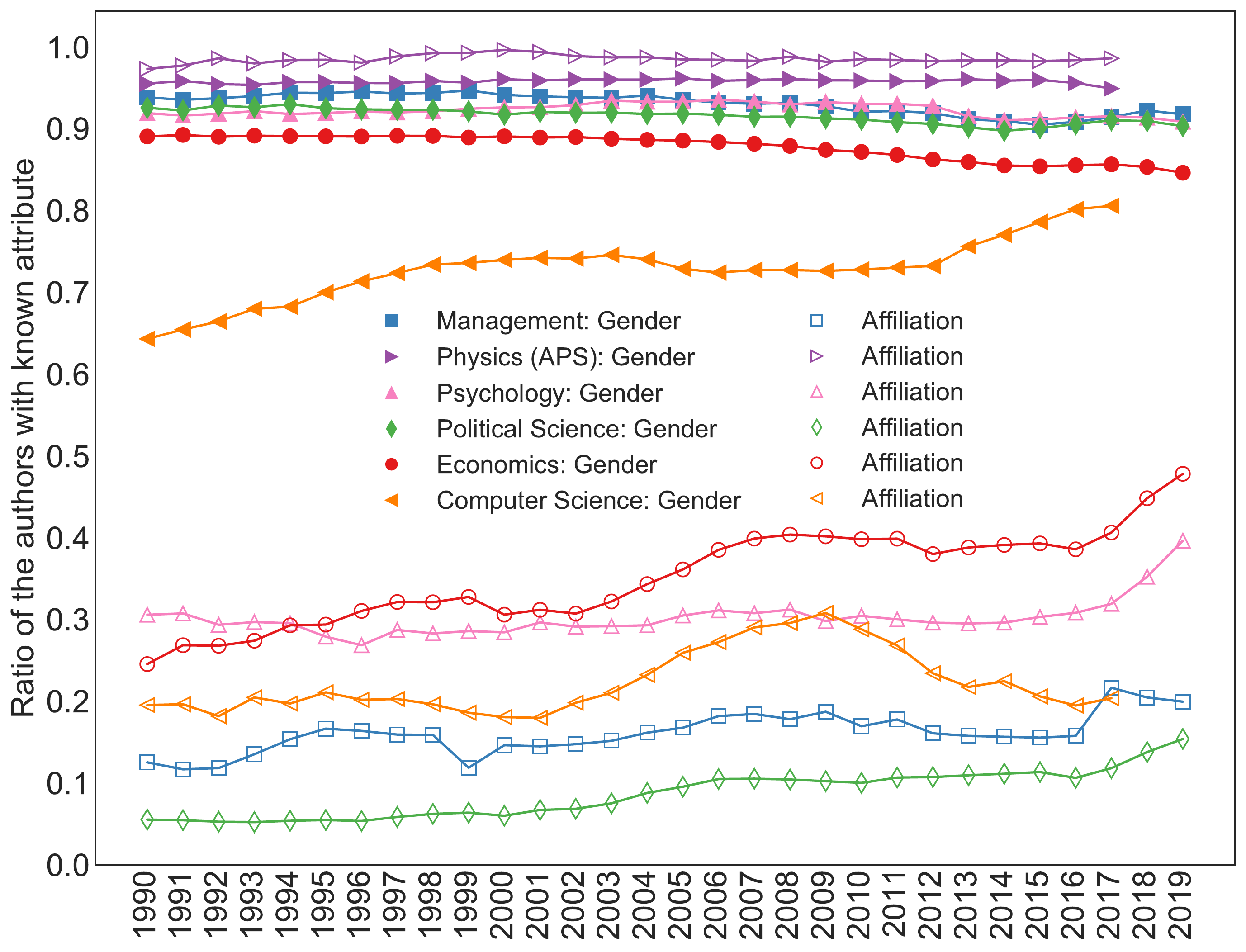}
    \caption{The fraction of authors with a known attribute (i.e.,~the coverage) over the years. The gender attribute has better coverage compared to affiliation of authors. We discard authors with unknown gender, and we consider authors with unknown affiliation in majority group (not affiliated with top-ranked universities). 
    }
    \label{fig:data_coverage}
\end{figure}
        

\subsection{Parameter Estimation Details}
\subsubsection{Citation Edge Ordering}
In our proposed DMPA model, citations edges form asynchronously and at a finer temporal scale than supported by the empirical data. To better align the model with the data, 
we assign an ordering to the edges so as to keep track of the new edges formed in the three events (citations to or from a new node, or between existing nodes). We create an ordered list of edges $E_{ord} = [(u_1, v_1), (u_2, v_2), ..., (u_n, v_n)]$ as follows. Starting with a \textit{seed} set with one author from the largest connected component of authors publishing in 1990, we traverse citations edges originating or terminating at these nodes and add them to the edge list $E_{ord}$ in random order. We then update the seed set by adding nodes that originate or terminate at the seed nodes and repeat the procedure until we cover all the years. The  algorithm is specified in more detail below: 
\begin{algorithm}[H]
\SetAlgoLined

\SetKwInOut{Input}{input}
\SetKwInOut{Output}{output}
\Input{graph $G$}
\Output{list of edge-ordering $E_{ord}$}

 construct graph $H$ using edges of $G$ in year $1990$\;
 
 pick one of the authors from the largest connected component of $H$ and name it as $v$\; 
 
 initialize seed $S = \{ v \}$ and $E_{ord} = []$\;
 
 \For{$y$ from $1990$ to $2020$}{
    \While{There is any potential edges to add from year $y$} {
        define $A$ as a list of edges which at least one of its sides are inside the seed set $S$\;
        
        shuffle list $A$\;
        
        append elements of $A$ to the end of $E_{ord}$\;
        
        update seed $S$ if there is a new node covered by $E_{ord}$\;
    }
 }
 \caption{Ordering of citation edges}
 \label{alg:ordering_citations}
\end{algorithm}

\noindent 
Note that the Algorithm~\ref{alg:ordering_citations} preserves the semi-dynamic ordering, meaning that there is no edge $a$ forming before edge $b$ in $E_{ord}$ where $a$ was created after $b$. Some of the edges in the citation graph $G$ will not appear in the edge list $E_{ord}$. However, since these edges are not part of the largest connected component of $G$, they likely connect to authors outside the field of study. 
Using this algorithm we cover at least $97\%$ of the edges forming during the period 1990--2020 in each field of study. We ran the parameter estimation for five times, and the standard deviation of the estimated parameters was less than $\num{1e-4}$ for all the parameters. This suggests that the randomization of the edges does not change the estimated parameters.

Next, we estimate model parameters using the new edge ordering. The number of edges generated by the arrival of new nodes (the first two types of citation events) is small compared to the number of edges generated between the existing nodes~(densification events) and therefore, empirical estimation of $\matrixEventOne$ and $\matrixEventTwo$ is not accurate due to lack of data points. 
As a result, we focus on densification events, where edges form between existing nodes, to estimate $\matrixEventThree$. We then assume that  $\matrixEventOne = \matrixEventTwo = \matrixEventThree = \matrixAnyEvent$ similar to Theorem~\ref{th:convergence_DMPA}. First, we estimated parameters $\redBirthProb$, $\probEventOne$, $\probEventTwo$, $\thetaIn$ and $\thetaOut$ using data as discussed in detail below. We then used a hyper-parameter tuning technique to estimate the preferential attachment parameter~$\deltaboth$. 
Finally, we used all the estimated parameters discussed so far in order to estimate
$\homophilyRedAnyEvent$ and $\homophilyBlueAnyEvent$ which constitute the elements of the matrix
$\matrixAnyEvent$.

\subsubsection{Estimating Class Balance Parameter $\redBirthProb$}
Recall that $\redBirthProb \in [0,1]$ represents the fraction of red nodes. 
We label authors by their gender or prestige of their institutional affiliation as described in the Methods section. For gender-partitioned networks, $\redBirthProb$ is the fraction of authors with female or unisex names. Note that this may overestimate the fraction of female authors. For affiliation-partitioned networks, $\redBirthProb$ is the fraction of authors from top-ranked institutions as defined by the Shanghai University Ranking.

\subsubsection{Estimating Edge Formation Rates $\probEventOne$, $\probEventTwo$}
For estimating the parameters $\probEventOne$ and $\probEventTwo$, we need to keep track of newly joined nodes over time. So, starting from empty set $S$, we iterate over all the edges in $E_{ord}$ and for each edge $(u, v)$ we add both $u$ and $v$ to set $S$ if they are not in the set already. Having $S$, parameter $\probEventOne$ (resp. $\probEventTwo$) could be estimated by counting the number of outgoing edges from (resp. incoming edges to) the newly added nodes to set $S$. In the other words, we need to count the number of times $u$ (cited node) was not in set $S$ and the number of times $v$ (citing node) was not in the set to estimate $p$ and $q$ respectively. 

\subsubsection{Estimating Preferential Attachment Parameter $\deltaboth$}
We find the best value of $\deltaboth$ (i.e.,~the value that results in the best set of model parameters in the sense of maximum likelihood)  through a hyper-parameter tuning approach as follows. We first estimate all parameters using different values of $\deltaboth \in \{1, 2, 3, 4, 5, 10, 20, 50, 100, 1000\}$ and use them to calculate the theoretical value of power-inequality $\powerInequality_{theoretical}$, if the condition given in  Eq.~\ref{eq:convergence}~(to ensure the convergence) is satisfied. We then select the parameter set that leads to power-inequality value $\powerInequality$ that is closest to its empirical value. 

\subsubsection{Estimating Homophily Parameters $\homophilyRedAnyEvent$, $\homophilyBlueAnyEvent$}
We estimate $\homophilyRedAnyEvent$ and $\homophilyBlueAnyEvent$ using the number of generated edges. Considering only the edges forming between existing nodes, we define $N_{rr}$ as the number of edges where both 
citing and cited nodes are red nodes, and $N_{xr}$ as number of edges where the citing node is red and the cited node could be red or blue. Then we can calculate $h_{\mathcal{R}} = \frac{N_{rr}}{N_{xr}}$ from the data. $\pthreeRR$ (given in Eq.~\ref{eq:pthreee_RR}) and $\pthreeBR$ (given in Eq.~\ref{eq:pthreee_BR}) are the probabilities of generating a red-to-red and blue-to-red edges in event 3, respectively. Using those expressions, we can write $h_\mathcal{R}$ as:

\begin{equation}
\frac{N_{rr}}{N_{xr}} = h_{\mathcal{R}} = \frac{\pthreeRR}{\pthreeRR + \pthreeBR} = \frac{a \times \homophilyRedAnyEvent}{a \times \homophilyRedAnyEvent + b \times (1 - \homophilyRedAnyEvent)}
\rightarrow \homophilyRedAnyEvent = \frac{b \times h_{\mathcal{R}}}{a - a \times h_{\mathcal{R}} + b \times h_{\mathcal{R}}}
\label{eq:rur_esimate}
\end{equation}

Here $a$ and $b$ are: 
\begin{equation}
a = \left(\thetaOut_\timeValue + \left(\probEventOne + \probEventTwo\right)\redBirthProb\deltaboth \right)\left(\thetaIn_\timeValue + \left(\probEventOne + \probEventTwo\right)\redBirthProb\deltaboth \right)
\end{equation}

\begin{equation}
b = \left(1-\thetaOut_\timeValue + \left(\probEventOne + \probEventTwo\right)\left(1-\redBirthProb\right)\deltaboth \right)\left(\thetaIn_\timeValue + \left(\probEventOne + \probEventTwo\right)\redBirthProb\deltaboth \right)
\end{equation}

Similarly, based on Eq.~\ref{eq:pthreee_BB} and Eq.~\ref{eq:pthreee_RB}, we have: 

\begin{equation}
\frac{N_{bb}}{N_{xb}} = h_{\mathcal{B}} = \frac{\pthreeBB}{\pthreeBB + \pthreeRB} = \frac{c \times \homophilyBlueAnyEvent}{c \times \homophilyBlueAnyEvent + d \times (1 - \homophilyBlueAnyEvent)}
\rightarrow \homophilyBlueAnyEvent = \frac{d \times h_{\mathcal{B}}}{c - c \times h_{\mathcal{B}}  + d \times h_{\mathcal{B}} },
\label{eq:rub_esimate}
\end{equation}

\noindent where $c$ and $d$ are

\begin{equation}
c = \left(1 - \thetaOut_\timeValue + \left(\probEventOne + \probEventTwo\right)\left(1-\redBirthProb\right)\deltaboth \right)\left(1-\thetaIn_\timeValue + \left(\probEventOne + \probEventTwo\right)\left(1-\redBirthProb\right)\deltaboth \right)
\end{equation}

\begin{equation}
d = \left(\thetaOut_\timeValue + \left(\probEventOne + \probEventTwo\right)\redBirthProb\deltaboth \right)\left(1-\thetaIn_\timeValue + \left(\probEventOne + \probEventTwo\right)\left(1-\redBirthProb\right)\deltaboth \right)
\end{equation}

\noindent Parameter $\thetaIn$ is the fraction of the total in-degree of red group and $\thetaOut$ is fraction of the total out-degree of red group, and could be estimated from data. Parameters $\redBirthProb$, $\probEventOne$ and $\probEventTwo$ and $\deltaboth$ could be estimated as described above. So, we can estimate $\homophilyRedAnyEvent$ and $\homophilyBlueAnyEvent$ using other parameters and equations \ref{eq:rur_esimate} and \ref{eq:rub_esimate}. 
Note that we are assuming citation networks have fixed parameters over time (i.e. $\thetaIn$, $\thetaOut$, $\redBirthProb$, ...). However, this is not the case in the citation network. The ratio of female/elite authors, for example, changes over time. This could explain the disparity between theoretical power-inequality and empirical power-inequality in Table~\ref{tab:params}.

\definecolor{Grey}{gray}{0.9}
\begin{table}[h!]
    \centering
    \begin{tabular}{|c|c|c|c|c|c|c|}
    \hline
    \rowcolor{Grey}
    \multicolumn{7}{|c|}{\textit{Gender-partitioned Network Parameters}} \\
        \hline 
        \rowcolor{Grey}
         & \textit{Management} & \textit{Physics (APS)} & \textit{Psychology} & \textit{Political Science} & \textit{Economics} & \textit{Computer Science} \\
        \hline 
        $T$ & $1.19M$ & $17.23M$ & $12.79M$ & $7.24M$ & $94.40M$ & $435.66M$\\ 
        \hline 
        $\deltaboth$ & $1000$ & $1000$ & $4$ & $1000$ & $100$ & $20$ \\ 
        \hline 
        $r$ & $0.35$ & $0.16$ & $0.50$  & $0.34$ & $0.28$ & $0.26$\\ 
        \hline 
        $p$ & $0.025$ & $0.001$ & $0.030$  & $0.067$ & $0.004$ & $0.005$\\ 
        \hline 
        $q$ & $0.058$ & $0.008$ & $0.032$  & $0.064$& $0.009$ & $0.012$ \\ 
        \hline 
        $\homophilyRedAnyEvent$ & $0.46$ & $0.48$ & $0.54$ & $0.47$ & $0.48$ & $0.55$ \\
        \hline 
        $\homophilyBlueAnyEvent$ & $0.61$ & $0.62$ & $0.57$ & $0.67$ & $0.62$ & $0.57$ \\
        \hline
        $\powerInequality_{\mathrm{empirical}}$ & $0.70$ & $0.76$ & $0.94$ & $0.61$ & $0.66$ & $0.80$ \\
        \specialrule{.12em}{.1em}{.1em}
        $\powerInequality_{\mathrm{theoretical}}$  & $0.74$ & $0.72$ & $0.82$ & $0.69$ & $0.65$ & $0.87$ \\ 
        \hline
        \hline
        \rowcolor{Grey}
        \multicolumn{7}{|c|}{\textit{Affiliation-partitioned Network Parameters}} \\
        \hline 
        \rowcolor{Grey}
         & \textit{Management} & \textit{Physics (APS)} & \textit{Psychology} & \textit{Political Science} & \textit{Economics} & \textit{Computer Science} \\
        \hline 
        $T$ & $842.85K$ & $15.25M$ & $12.61M$ & $2.61M$ & $91.51M$ & $419.24M$ \\ 
        \hline 
        $\deltaboth$ & $1000$ & $1000$ & $1000$ & $10$ & $1000$ & $1000$ \\ 
        \hline 
        $r$ & $0.12$ & $0.22$ & $0.24$ & $0.18$ & $0.13$ & $0.09$ \\ 
        \hline 
        $p$ & $0.012$ & $0.001$ & $0.024$ & $0.020$ & $0.001$ & $0.001$\\ 
        \hline 
        $q$ & $0.048$ & $0.009$ & $0.026$ & $0.053$ & $0.006$ & $0.007$\\ 
        \hline 
        $\homophilyRedAnyEvent$ & $0.83$ & $0.72$ & $0.74$ & $0.72$ & $0.81$ & $0.85$ \\
        \hline 
        $\homophilyBlueAnyEvent$ & $0.29$ & $0.44$ & $0.47$ & $0.44$ & $0.30$ & $0.31$ \\
        \hline 
        $\powerInequality_{\mathrm{empirical}}$ & $3.11$ & $1.52$ & $1.11$ & $1.86$ & $2.31$ & $1.91$ \\
        \specialrule{.12em}{.1em}{.1em}
        $\powerInequality_{\mathrm{theoretical}}$  & $3.53$ & $1.94$ & $1.79$ & $3.32$ & $3.89$ & $4.54$ \\ 
        \hline 
    \end{tabular}
    \caption{Estimated values of the parameters of DMPA model. These values are also illustrated visually in Fig.~\ref{fig:estimated_parameters}. Parameters for different fields of study
    are estimated from Microsoft Academic Graph data, except for Physics, which is estimated from data provided by the American Physical Society (APS). The parameters are estimated using the procedure described in Supplementary Information. These estimated parameter values together with the insights provided by the DMPA model allow us to develop strategies for mitigating power-inequality in science as discussed in Results section.}
    \label{tab:params}
\end{table}

\subsection{Proof of Theorem~\ref{th:convergence_DMPA}}

\subsubsection{Preliminaries and the outline of the proof}
The proof of Theorem~\ref{th:convergence_DMPA} uses the following result from~\cite[Theorem~2.1]{kushner2003}.
\begin{theorem}[Convergence w.p.~1 under Martingale difference noise~\cite{kushner2003}]
	\label{th:kushner_theorem}
	Consider the algorithm
	\begin{equation}
	\label{eq:algorithm_kushner_theorem}
	x_{\timeValue + 1} = x_{\timeValue} + \epsilon_\timeValue Y_\timeValue +\epsilon_\timeValue z_\timeValue
	\end{equation}
where,	$Y_\timeValue \in \mathbb{R}^m$, $\epsilon_\timeValue$ denotes a decreasing step size and $\epsilon_\timeValue z_\timeValue$ is the shortest Euclidean length needed to project $x_\timeValue$ into some compact set $H$. Assume,
	\begin{enumerate}[leftmargin = 0.5in]
		\item[(C.1)]  $\sup_\timeValue |Y_\timeValue|^2 < \infty$
		
		\item[(C.2)] There exists a measurable function $\bar{g}(\cdot)$ of $x$ and random variables $\beta_\timeValue$ such that, 
		\begin{equation}
		\mathbb{E}_\timeValue Y_\timeValue = \mathbb{E}\{Y_\timeValue|x_0, Y_i, i < \timeValue\} = \bar{g}(x_\timeValue) +\beta_\timeValue
		\end{equation}
		
		\item[(C.3)] $\bar{g}(\cdot)$ is Lipschitz continuous
		
		\item[(C.4)] $\epsilon_\timeValue \geq 0, \epsilon_i \rightarrow 0$ for $i \geq 0$ and $\epsilon_{\timeValue} = 0$ for $i < 0$
		
		\item[(C.5)] $\sum_{\timeValue}\epsilon_{\timeValue}^2 < \infty, \sum_{i}\epsilon_i = \infty$
		
		\item[(C.6)] $\sum_{i}\epsilon_i\left|\beta_i\right| < \infty$	w.p.~1
	\end{enumerate}
If $\bar{x}$ is an asymptotically stable point of the ordinary differential equation~(ODE)
\begin{equation}
\label{eq:ODE_kushner_theorem}
 \dot{x} = \bar{g}(x) + z
\end{equation} where $z$ is the shortest Euclidean length needed to ensure $x$ is in $H$ and, $x_\timeValue$ is in some compact set in the domain of attraction of $\bar{x}$ infinitely often with probability $\geq \rho$, then $x_\timeValue \rightarrow \bar{x}$ with at least probability~$\rho$. 
\end{theorem} 

In the context of Theorem~\ref{th:kushner_theorem}, the idea behind the proof of Theorem~\ref{th:convergence_DMPA} is to express the evolution of~$\thetaVec_\timeValue$ in the form of Eq.~\ref{eq:algorithm_kushner_theorem} with $z_\timeValue = 0$ and, show that the assumptions C.1-C.6 are satisfied. Then we show that the function~$\bar{g}$ is of the form $\bar{g}\left(\thetaVec_\timeValue\right) = \nonLinFunction(\thetaVec_\timeValue) - \thetaVec_\timeValue$ where $\nonLinFunction(\cdot)$ is a contraction map. Theorem~\ref{th:kushner_theorem} thus implies that there exists a globally asymptotically stable equilibrium state~(which is the unique fixed point of the contraction map~$\nonLinFunction(\cdot)$) to which the sequence $\{\thetaVec_\timeValue\}$ converges almost surely. 

Thus, in fact, Theorem 1 specifies more than the asymptotic normalized in- and out--degrees. It also implies that normalized in and out-degrees~(i.e.,~$\thetaVec_\timeValue$) evolving as a random process over time (appropriately scaled) converge in sample path to a deterministic ODE.

\subsubsection{Proof of Theorem~\ref{th:convergence_DMPA}}
{\bf Notation:} We first need some additional notation that for the proof of first part. Let,
\begin{itemize}
		\item[] $\poneBB : $ given the event~$1$ and the new node is blue, probability that it is followed by an existing blue node
		\item[] $\poneBR: $ given the event~$1$ and the new node is blue, probability that it is followed by an existing red node
		\item[] $\poneRR : $ given the event~$1$  and the new node is red, probability that  it is followed by an existing red node
		\item[] $\poneRB : $ given the event~$1$ and the new node is red, probability that it is followed by an existing blue node.
\end{itemize}
Analogous quantities for event~2 and event~3 are denoted by $\ptwoBB$, $\ptwoBR$, $\ptwoRR$, $\ptwoRB$ and $\pthreeBB$, $\pthreeBR$, $\pthreeRR$, $\pthreeRB$, respectively. Further, let $n(\calRed_\timeValue), n(\calBlue_\timeValue)$ denote the number of nodes at time $\timeValue$ in the red and blue groups respectively and $\groupSize_\timeValue
= n(\calRed_\timeValue) + n(\calBlue_\timeValue)$.

\vspace{0.2cm}
With the above notation, we can express the evolution of the total in-degree of red nodes~$\inDegree(\calRed_{\timeValue})$ as follows:
\begin{align}
\label{eq:proof_degree_i_evolution}
\mathbb{E}\{\inDegree({\calRed_{\timeValue + 1})} - \inDegree({\calRed_{\timeValue})} |
\graph_\timeValue \} &= \probEventOne(\redBirthProb\poneRR + (1-\redBirthProb)\poneBR) + \probEventTwo\redBirthProb + (1-\probEventOne - \probEventTwo)(\pthreeRR + \pthreeBR).
\end{align}
The idea behind Eq.~\ref{eq:proof_degree_i_evolution} is that the total in-degree of red group can increase by an amount of $1$ at time $\timeValue$~(i.e.,~$\inDegree({\calRed_{\timeValue + 1})} - \inDegree({\calRed_{\timeValue})} = 1$) via one of three mutually exclusive ways: event~1 takes place and either a new red node is added and follows an existing red node or a new blue node is added and follows an existing red node, event~2 takes place and a new red node is born, or event~3 takes place and an an existing red node follows an existing blue or a red node. Eq.~\ref{eq:proof_degree_i_evolution} expresses the expectation of this event where the three summands correspond to the three ways in which the in-degree of the red-group increase by one. Since the number of edges in the network $\graph_\timeValue$ at time $\timeValue$ is equal to $\timeValue$~(i.e.,~$|\edgeSet_\timeValue| = \timeValue$), from Eq.~\ref{eq:proof_degree_i_evolution} we get,
\begin{align}
&\mathbb{E}\left\{\frac{\inDegree\left(\calRed_{\timeValue + 1}\right)}{\timeValue + 1}|
\graph_\timeValue \right\} =\mathbb{E}\left\{{\thetaIn_{\timeValue + 1}}|
\graph_\timeValue \right\} \nonumber\\ &\hspace{0.5cm}=\frac{\inDegree({\calRed_{\timeValue})}}{\timeValue + 1} +  \frac{1}{\timeValue + 1}\left(\probEventOne(\redBirthProb\poneRR + (1-\redBirthProb)\poneBR) + \probEventTwo\redBirthProb + (1-\probEventOne - \probEventTwo)(\pthreeRR + \pthreeBR) \right) \nonumber\\
&\hspace{0.5cm}= \thetaIn_\timeValue + \frac{1}{\timeValue + 1}\left(\probEventOne(\redBirthProb\poneRR + (1-\redBirthProb)\poneBR) + \probEventTwo\redBirthProb + (1-\probEventOne - \probEventTwo)(\pthreeRR + \pthreeBR) -\thetaIn_\timeValue\right). \label{eq:proof_theta_i_evolution}
\end{align}

Using similar arguments for the total out-degree of red group, we get,
\begin{align}
&\mathbb{E}\left\{\frac{\outDegree\left(\calRed_{\timeValue + 1}\right)}{\timeValue + 1}|
\graph_\timeValue \right\}  = \mathbb{E}\left\{{\thetaOut_{\timeValue + 1}}|
\graph_\timeValue \right\} \nonumber\\ 
&\hspace{0.5cm}= \thetaOut_\timeValue + \frac{1}{\timeValue + 1}\left( \probEventOne\redBirthProb + \probEventTwo(\redBirthProb\ptwoRR + (1-\redBirthProb)\ptwoBR) + (1-\probEventOne - \probEventTwo)(\pthreeRR + \pthreeRB) -\thetaOut_\timeValue\right). \label{eq:proof_theta_o_evolution}
\end{align}

Therefore, by combining Eq.~\ref{eq:proof_theta_i_evolution} and Eq.~\ref{eq:proof_theta_o_evolution}, we get,
\begin{align}
&\mathbb{E}\left\{{\thetaVec_{\timeValue + 1}}| \graph_\timeValue\right\}  = \thetaVec_\timeValue + \nonumber\\
&\hspace{0.25cm}\frac{1}{\timeValue + 1}
\begin{bmatrix}
\probEventOne(\redBirthProb\poneRR + (1-\redBirthProb)\poneBR) + \probEventTwo\redBirthProb + (1-\probEventOne - \probEventTwo)(\pthreeRR + \pthreeBR) -\thetaIn_\timeValue  \\
 \probEventOne\redBirthProb + \probEventTwo(\redBirthProb\ptwoRR + (1-\redBirthProb)\ptwoBR) + (1-\probEventOne - \probEventTwo)(\pthreeRR + \pthreeRB) -\thetaOut_\timeValue \label{eq:proof_theta_evolution}
\end{bmatrix}.
\end{align}

Next, let us consider~$\poneRR$. Recall that $\poneRR$ is the probability that, given the event~1 happened and the new node is red, it is followed by an existing red node. This event can happen in three ways: \\
 1~-~a potential red follower is chosen via preferential attachment~(with probability~$\frac{\inDegree(\calRed_{\timeValue}) + \groupSize(\calRed_\timeValue)\deltaboth}{\degree_\timeValue + \groupSize_\timeValue \deltaboth}$) and the new node is followed by it~(probability $\homophilyRedAnyEvent$)
  \\2~-~a potential red follower is chosen via preferential attachment~(with probability~$\frac{\inDegree(\calRed_{\timeValue}) + \groupSize(\calRed_\timeValue)\deltaboth}{\degree_\timeValue + \groupSize_\timeValue \deltaboth}$) and the new node is not followed by it~(probability $1-\homophilyRedAnyEvent$) and then the event takes place after that with probability $\poneRR$\\
 3~-~a potential blue follower is chosen via preferential attachment~(with probability~$\frac{\inDegree(\calBlue_{\timeValue}) + \groupSize(\calBlue_\timeValue)\deltaboth}{\degree_\timeValue + \groupSize_\timeValue \deltaboth}$) and the new node is not followed by it~(probability $\homophilyBlueAnyEvent$) and then the event takes place after that with probability $\poneRR$.

 Hence, $\poneRR$ satisfies,
\begin{align}
\poneRR &= \frac{\inDegree(\calRed_{\timeValue}) + \groupSize(\calRed_\timeValue)}{\degree_\timeValue + \groupSize_\timeValue \deltaboth}\homophilyRedAnyEvent + \left(\frac{\inDegree(\calBlue_{\timeValue}) + \groupSize(\calBlue_\timeValue)\deltaboth}{\degree_\timeValue + \groupSize_\timeValue \deltaboth}\homophilyBlueAnyEvent + \frac{\inDegree(\calRed_{\timeValue}) + \groupSize(\calRed_\timeValue)\deltaboth}{\degree_\timeValue + \groupSize_\timeValue \deltaboth}\left(1-\homophilyRedAnyEvent\right) \right)\poneRR \nonumber\\
 &= \frac{ \left(\inDegree(\calRed_\timeValue) + \groupSize(\calRed_{\timeValue})\deltaboth\right)\homophilyRedAnyEvent}{\degree_\timeValue + \groupSize_\timeValue \deltaboth -\left(		 \left(\inDegree(\calBlue_\timeValue) + \groupSize(\calBlue_{\timeValue})\deltaboth\right)\homophilyBlueAnyEvent	+  \left(\inDegree(\calRed_\timeValue) + \groupSize(\calRed_{\timeValue})\deltaboth\right)\left(1-\homophilyRedAnyEvent\right)\right)} \nonumber\\
 &= \frac{\left(\thetaIn_\timeValue\degree_\timeValue + \groupSize(\calRed_\timeValue)\deltaboth \right)\homophilyRedAnyEvent}{\degree_\timeValue + \groupSize_\timeValue \deltaboth -		 \left( \left(1-\thetaIn_\timeValue\right) \degree_\timeValue + \groupSize(\calBlue_{\timeValue})\deltaboth \right)\homophilyBlueAnyEvent	 -\left( \thetaIn_\timeValue \degree_\timeValue+ \groupSize(\calRed_{\timeValue})\deltaboth\right)\left(1-\homophilyRedAnyEvent\right)} \quad \text{(by definition of $\thetaIn_\timeValue$)} \nonumber\\
 &= \frac{\left(\thetaIn_\timeValue\timeValue + \groupSize(\calRed_\timeValue)\deltaboth \right)\homophilyRedAnyEvent}{\timeValue + \groupSize_\timeValue \deltaboth -		 \left( \left(1-\thetaIn_\timeValue\right) \timeValue + \groupSize(\calBlue_{\timeValue})\deltaboth \right)\homophilyBlueAnyEvent	 -\left( \thetaIn_\timeValue \timeValue+ \groupSize(\calRed_{\timeValue})\deltaboth\right)\left(1-\homophilyRedAnyEvent\right)} \quad \text{(since $\degree_\timeValue = \timeValue$)} \label{eq:poneRR}
\end{align}
Next, by Hoeffding's inequality, we get,
\begin{equation}
\label{eq:hoeffding_poneRR}
\begin{aligned}
\mathbb{P}\left\{	\left|\groupSize\left(\calRed_{\timeValue} \right) -(\probEventOne + \probEventTwo)\redBirthProb\timeValue \right|	\geq \sqrt{\frac{\timeValue \log \timeValue}{2}}\right\} &\leq \frac{1}{\timeValue}\\
\mathbb{P}\left\{	\left|\groupSize\left(\calBlue_{\timeValue} \right) -(\probEventOne + \probEventTwo)\left(1-\redBirthProb\right)\timeValue \right|	\geq \sqrt{\frac{\timeValue \log \timeValue}{2}}\right\} &\leq \frac{1}{\timeValue}\\
\mathbb{P}\left\{	\left|\groupSize_\timeValue -(\probEventOne + \probEventTwo)\timeValue \right|	\geq \sqrt{\frac{\timeValue \log \timeValue}{2}}\right\} &\leq \frac{1}{\timeValue}.
\end{aligned}
\end{equation}
Hence, by Eq.~\ref{eq:poneRR} and Eq.~\ref{eq:hoeffding_poneRR}, we observe that, with probability $1-o(\frac{1}{\timeValue})$,
\begin{align}
\poneRR&= \frac{\left(\thetaIn_\timeValue + \left(\probEventOne+\probEventTwo\right)\redBirthProb\deltaboth \right)\homophilyRedAnyEvent}{1 + \left(\probEventOne+\probEventTwo\right)\deltaboth -		 \left( \left(1-\thetaIn_\timeValue\right) + \left(\probEventOne+\probEventTwo\right)\left(1-\redBirthProb\right)\deltaboth \right)\homophilyBlueAnyEvent	 -\left( \thetaIn_\timeValue+ \left(\probEventOne+\probEventTwo\right)\redBirthProb\deltaboth\right)\left(1-\homophilyRedAnyEvent\right)} + O\left(\frac{1}{\timeValue^{\frac{1}{4}}}\right). \nonumber\\
\end{align}
Following similar steps, we also get, with probability $1-o(\frac{1}{\timeValue})$,
\begin{align}
\poneBR&= \frac{\left(\thetaIn_\timeValue + \left(\probEventOne+\probEventTwo\right)\redBirthProb\deltaboth \right)\left(1-\homophilyRedAnyEvent\right)}{1 + \left(\probEventOne+\probEventTwo\right)\deltaboth -		 \left( \left(1-\thetaIn_\timeValue\right) + \left(\probEventOne+\probEventTwo\right)\left(1-\redBirthProb\right)\deltaboth \right)\left(1-\homophilyBlueAnyEvent\right) -\left( \thetaIn_\timeValue+ \left(\probEventOne+\probEventTwo\right)\redBirthProb\deltaboth\right)\homophilyRedAnyEvent} + O\left(\frac{1}{\timeValue^{\frac{1}{4}}}\right). \nonumber\\
\ptwoRR&= \frac{\left(\thetaOut_\timeValue + \left(\probEventOne+\probEventTwo\right)\redBirthProb\deltaboth \right)\homophilyRedAnyEvent}{1 + \left(\probEventOne+\probEventTwo\right)\deltaboth -		 \left( \left(1-\thetaOut_\timeValue\right) + \left(\probEventOne+\probEventTwo\right)\left(1-\redBirthProb\right)\deltaboth \right)\homophilyRedAnyEvent	 -\left( \thetaOut_\timeValue+ \left(\probEventOne+\probEventTwo\right)\redBirthProb\deltaboth\right)\left(1-\homophilyRedAnyEvent\right)} + O\left(\frac{1}{\timeValue^{\frac{1}{4}}}\right). \nonumber\\
\ptwoBR&= \frac{\left(\thetaOut_\timeValue + \left(\probEventOne+\probEventTwo\right)\redBirthProb\deltaboth \right)\left(1-\homophilyBlueAnyEvent\right)}{1 + \left(\probEventOne+\probEventTwo\right)\deltaboth -		 \left( \left(1-\thetaOut_\timeValue\right) + \left(\probEventOne+\probEventTwo\right)\left(1-\redBirthProb\right)\deltaboth \right)\left(1-\homophilyBlueAnyEvent\right)	 -\left( \thetaOut_\timeValue+ \left(\probEventOne+\probEventTwo\right)\redBirthProb\deltaboth\right)\homophilyBlueAnyEvent} + O\left(\frac{1}{\timeValue^{\frac{1}{4}}}\right). \nonumber
\end{align} and,
\begin{align}
\pthreeRR&= \frac{\left(\thetaOut_\timeValue + \left(\probEventOne + \probEventTwo\right)\redBirthProb\deltaboth \right)\left(\thetaIn_\timeValue + \left(\probEventOne + \probEventTwo\right)\redBirthProb\deltaboth \right)\homophilyRedAnyEvent}{
	\splitfrac{\left(1+\left(\probEventOne+\probEventTwo\right)\deltaboth\right)^2 }{
	\splitfrac{
	 - \left( 1	-\thetaOut_\timeValue+ \left(\probEventOne+\probEventTwo\right)\left(1-\redBirthProb\right)\deltaboth\right)\left( 1	-\thetaIn_\timeValue+ \left(\probEventOne+\probEventTwo\right)\left(1-\redBirthProb\right)\deltaboth\right)\left(1-\homophilyBlueAnyEvent\right)}{\splitfrac{- \left( 1	-\thetaOut_\timeValue+ \left(\probEventOne+\probEventTwo\right)\left(1-\redBirthProb\right)\deltaboth\right)\left(\thetaIn_\timeValue+ \left(\probEventOne+\probEventTwo\right)\redBirthProb\deltaboth\right)\homophilyRedAnyEvent }{\splitfrac{- \left( \thetaOut_\timeValue+ \left(\probEventOne+\probEventTwo\right)\redBirthProb\deltaboth\right)\left( 1	-\thetaIn_\timeValue+ \left(\probEventOne+\probEventTwo\right)\left(1-\redBirthProb\right)\deltaboth\right)\homophilyBlueAnyEvent}{
	 	- \left(\thetaOut_\timeValue+ \left(\probEventOne+\probEventTwo\right)\redBirthProb\deltaboth\right)\left(	\thetaIn_\timeValue+ \left(\probEventOne+\probEventTwo\right)\redBirthProb\deltaboth\right)\left(1-\homophilyRedAnyEvent\right)
 	}	 
 } 
}
}
}	+ O\left(\frac{1}{\timeValue^{\frac{1}{4}}}\right)	\label{eq:pthreee_RR}\\
\pthreeBR&= \frac{\left(1-\thetaOut_\timeValue + \left(\probEventOne + \probEventTwo\right)\left(1-\redBirthProb\right)\deltaboth \right)\left(\thetaIn_\timeValue + \left(\probEventOne + \probEventTwo\right)\redBirthProb\deltaboth \right)\left(1-\homophilyRedAnyEvent\right)}{
	\splitfrac{\left(1+\left(\probEventOne+\probEventTwo\right)\deltaboth\right)^2 }{
		\splitfrac{
			- \left( 1	-\thetaOut_\timeValue+ \left(\probEventOne+\probEventTwo\right)\left(1-\redBirthProb\right)\deltaboth\right)\left( 1	-\thetaIn_\timeValue+ \left(\probEventOne+\probEventTwo\right)\left(1-\redBirthProb\right)\deltaboth\right)\left(1-\homophilyBlueAnyEvent\right)}{\splitfrac{- \left( 1	-\thetaOut_\timeValue+ \left(\probEventOne+\probEventTwo\right)\left(1-\redBirthProb\right)\deltaboth\right)\left(\thetaIn_\timeValue+ \left(\probEventOne+\probEventTwo\right)\redBirthProb\deltaboth\right)\homophilyRedAnyEvent }{\splitfrac{- \left( \thetaOut_\timeValue+ \left(\probEventOne+\probEventTwo\right)\redBirthProb\deltaboth\right)\left( 1	-\thetaIn_\timeValue+ \left(\probEventOne+\probEventTwo\right)\left(1-\redBirthProb\right)\deltaboth\right)\homophilyBlueAnyEvent}{
					- \left(\thetaOut_\timeValue+ \left(\probEventOne+\probEventTwo\right)\redBirthProb\deltaboth\right)\left(	\thetaIn_\timeValue+ \left(\probEventOne+\probEventTwo\right)\redBirthProb\deltaboth\right)\left(1-\homophilyRedAnyEvent\right)
				}	 
			} 
		}
	}
}	+ O\left(\frac{1}{\timeValue^{\frac{1}{4}}}\right)	\label{eq:pthreee_BR}\\
\pthreeRB&= \frac{\left(\thetaOut_\timeValue + \left(\probEventOne + \probEventTwo\right)\redBirthProb\deltaboth \right)\left(1-\thetaIn_\timeValue + \left(\probEventOne + \probEventTwo\right)\left(1-\redBirthProb\right)\deltaboth \right)\left(1-\homophilyBlueAnyEvent\right)}{
	\splitfrac{\left(1+\left(\probEventOne+\probEventTwo\right)\deltaboth\right)^2 }{
		\splitfrac{
			- \left( 1	-\thetaOut_\timeValue+ \left(\probEventOne+\probEventTwo\right)\left(1-\redBirthProb\right)\deltaboth\right)\left( 1	-\thetaIn_\timeValue+ \left(\probEventOne+\probEventTwo\right)\left(1-\redBirthProb\right)\deltaboth\right)\left(1-\homophilyBlueAnyEvent\right)}{\splitfrac{- \left( 1	-\thetaOut_\timeValue+ \left(\probEventOne+\probEventTwo\right)\left(1-\redBirthProb\right)\deltaboth\right)\left(\thetaIn_\timeValue+ \left(\probEventOne+\probEventTwo\right)\redBirthProb\deltaboth\right)\homophilyRedAnyEvent }{\splitfrac{- \left( \thetaOut_\timeValue+ \left(\probEventOne+\probEventTwo\right)\redBirthProb\deltaboth\right)\left( 1	-\thetaIn_\timeValue+ \left(\probEventOne+\probEventTwo\right)\left(1-\redBirthProb\right)\deltaboth\right)\homophilyBlueAnyEvent}{
					- \left(\thetaOut_\timeValue+ \left(\probEventOne+\probEventTwo\right)\redBirthProb\deltaboth\right)\left(	\thetaIn_\timeValue+ \left(\probEventOne+\probEventTwo\right)\redBirthProb\deltaboth\right)\left(1-\homophilyRedAnyEvent\right)
				}	 
			} 
		}
	}
}	+ O\left(\frac{1}{\timeValue^{\frac{1}{4}}}\right)	\label{eq:pthreee_RB}\\
\pthreeBB&= \frac{\left(1-\thetaOut_\timeValue + \left(\probEventOne + \probEventTwo\right)\left(1-\redBirthProb\right)\deltaboth \right)\left(1-\thetaIn_\timeValue + \left(\probEventOne + \probEventTwo\right)\left(1-\redBirthProb\right)\deltaboth \right)\homophilyBlueAnyEvent}{
	\splitfrac{\left(1+\left(\probEventOne+\probEventTwo\right)\deltaboth\right)^2 }{
		\splitfrac{
			- \left( 1	-\thetaOut_\timeValue+ \left(\probEventOne+\probEventTwo\right)\left(1-\redBirthProb\right)\deltaboth\right)\left( 1	-\thetaIn_\timeValue+ \left(\probEventOne+\probEventTwo\right)\left(1-\redBirthProb\right)\deltaboth\right)\left(1-\homophilyBlueAnyEvent\right)}{\splitfrac{- \left( 1	-\thetaOut_\timeValue+ \left(\probEventOne+\probEventTwo\right)\left(1-\redBirthProb\right)\deltaboth\right)\left(\thetaIn_\timeValue+ \left(\probEventOne+\probEventTwo\right)\redBirthProb\deltaboth\right)\homophilyRedAnyEvent }{\splitfrac{- \left( \thetaOut_\timeValue+ \left(\probEventOne+\probEventTwo\right)\redBirthProb\deltaboth\right)\left( 1	-\thetaIn_\timeValue+ \left(\probEventOne+\probEventTwo\right)\left(1-\redBirthProb\right)\deltaboth\right)\homophilyBlueAnyEvent}{
					- \left(\thetaOut_\timeValue+ \left(\probEventOne+\probEventTwo\right)\redBirthProb\deltaboth\right)\left(	\thetaIn_\timeValue+ \left(\probEventOne+\probEventTwo\right)\redBirthProb\deltaboth\right)\left(1-\homophilyRedAnyEvent\right)
				}	 
			} 
		}
	}
}	+ O\left(\frac{1}{\timeValue^{\frac{1}{4}}}\right)	\label{eq:pthreee_BB}
\end{align}

Now, we have all the ingredients to show that the assumptions C.1-C.6 are satisfied Theorem~\ref{th:kushner_theorem} for the sequence~$\{\thetaVec_\timeValue\}$. 

First, note that $\{\thetaVec_\timeValue\}$ is Markovian implying that its evolution can be expressed in the form Eq.~\ref{eq:algorithm_kushner_theorem} with $z_\timeValue = 0$ for all $\timeValue = 1, 2, \dots$ since $\thetaVec_\timeValue$ is naturally in the space $[0,1]^2$~i.e.,~$\thetaVec_{\timeValue + 1} = \thetaVec_\timeValue + \epsilon_{\timeValue}Y_\timeValue$. This further implies that C.1 is satisfied. 

Next, note that some of the terms on the right side of Eq.~\ref{eq:proof_theta_evolution} are constant model parameters~($\redBirthProb$, $\probEventOne$, $\probEventTwo$, $\homophilyBlueAnyEvent$, $\homophilyRedAnyEvent$, $\deltaboth$) and each of the remaining terms~($\poneRR$, $\poneBR$, $\ptwoRR$, $\ptwoBR$, $\pthreeRR$, $\pthreeBR$, $\pthreeRB$) is equal to a function of $\thetaIn_\timeValue, \thetaOut_\timeValue$ plus a noise term that is $O\left(\frac{1}{\timeValue^{{1}/{4}}}\right)$ with probability $1-o\left(\frac{1}{\timeValue}\right)$. Further, we can see from Eq.~\ref{eq:proof_theta_evolution} that $\epsilon_{\timeValue} = \frac{1}{\timeValue + 1}$ and since the noise term is $O\left(\frac{1}{\timeValue^{{1}/{4}}}\right)$, C.6 is satisfied. Further, we get,
\begin{equation}
\label{eq:proof_F_expression}
\begin{aligned}
g(\thetaVec_{\timeValue})&=\nonLinFunction(\thetaVec_\timeValue) - \thetaVec_{\timeValue}\\
\nonLinFunction(\thetaVec_\timeValue) &= \begin{bmatrix}
\probEventOne(\redBirthProb\barponeRR + (1-\redBirthProb)\barponeBR) + \probEventTwo\redBirthProb + (1-\probEventOne - \probEventTwo)(\barpthreeRR + \barpthreeBR) \\
\probEventOne\redBirthProb + \probEventTwo(\redBirthProb\barptwoRR + (1-\redBirthProb)\barptwoBR) + (1-\probEventOne - \probEventTwo)(\barpthreeRR + \barpthreeRB) 
\end{bmatrix}
\end{aligned}
\end{equation}
where, $\barponeRR$, $\barponeBR$, $\barptwoRR$, $\barptwoBR$, $\barpthreeRR$, $\barpthreeBR$, $\barpthreeRB$ correspond to $\poneRR$, $\poneBR$, $\ptwoRR$, $\ptwoBR$, $\pthreeRR$, $\pthreeBR$, $\pthreeRB$, respectively, with the $O\left(\frac{1}{\timeValue^{{1}/{4}}}\right)$ noise term neglected. 

Hence, the only remaining task left is to show that $\nonLinFunction(\cdot)$ (defined in Eq.~\ref{eq:proof_F_expression}) is a contraction map. For this, after some tedious algebra, we obtain the Jacobian $J$ of $\nonLinFunction(\cdot)$ to be of a special form with respect to $\deltaboth$: each element (denoted by~$J_{ij}\, i,j = 1,2$) of $J$ is a ratio of two polynomials of $\deltaboth$ where the denominator polynomial is the higher order one. This implies that there exists $\deltaboth^{*} >0$ such that, for all $\deltaboth > \deltaboth^{*}$,
\begin{align}
\left|\left|J\right|\right|_2 \leq \left|\left|J\right|\right|_F = \left(\sum_{i}\sum_{j}\left|J_{ij}\right|^2\right)^\frac{1}{2} < 1
\label{eq:convergence} 
\end{align}
where $\left|\left|\cdot\right|\right|_F$ denotes the Frobenius norm and $\left|\left|\cdot\right|\right|_2$ denotes the  largest singular value~(spectral norm). Hence, there exists $\deltaboth^{*} >0$ such that, for all $\deltaboth > \deltaboth^{*}$, $\nonLinFunction(\cdot)$ is a contraction map. 
 
\subsection{Additional Details of the DMPA model}
\begin{figure}
	\includegraphics[width=\columnwidth,  trim={0.15cm 0.3cm 0.2cm 0.1cm},clip]{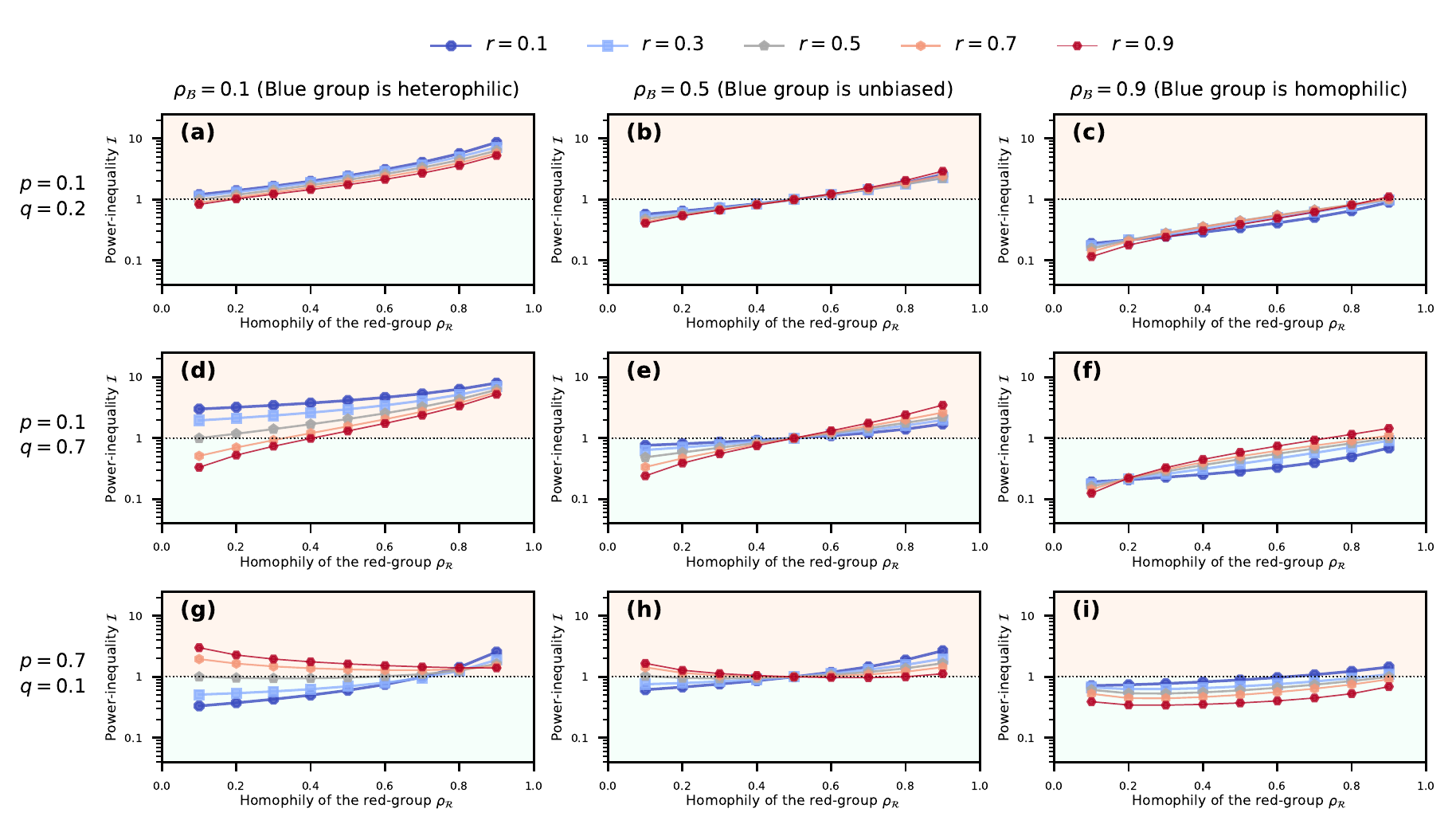}
	\caption{
The aim of this figure is to illustrate how preferential-attachment affects power-inequality. The plots show the variation of the asymptotic power-inequality~$\powerInequality$ with the homophily of the red group~$\homophilyRedAnyEvent$ under the DMPA model with $\deltaboth = 100$ analogous to Fig.~\ref{fig:PowerInequalityTheoretical} (where $\deltaboth = 10$).
Comparing this figure with Fig.~\ref{fig:PowerInequalityTheoretical} shows that increasing $\deltaboth$~(i.e.,~reducing the preferential attachment) reduces the effect of the group size on power-inequality, especially when one group is highly homophilic and the other heterophilic. Thus, reducing preferential-attachment is an effective strategy to ameliorate power-inequality in the presence of extreme class-imbalance and differences in homophily parameters of the two groups.
} 
	\label{fig:PowerInequalityTheoretical_delta100}
\end{figure}

Fig.~\ref{fig:PowerInequalityTheoretical_delta100} illustrates the asymptotic power-inequality $\powerInequality$ under the DMPA model for different parameter combinations with fixed $\deltaboth = 100$. Comparing Fig.~\ref{fig:PowerInequalityTheoretical}~(generated with $\deltaboth = 10$) with Fig.~\ref{fig:PowerInequalityTheoretical_delta100} shows that increasing $\deltaboth$ (i.e.,~reducing preferential-attachment) reduces the power-inequality. In particular, the effect of $\deltaboth$ on power-inequality is larger when there is a high degree of class imbalance (i.e.,~$\redBirthProb \ll 0.5$ or $\redBirthProb \gg 0.5$).

\subsubsection*{Code and details}
The codes for empirical and simulation results as well as the complete datasets are available on GitHub: \url{https://github.com/ninoch/DMPA}


\acknow{
B.~Nettasinghe and V.~Krishnamurthy are funded in part by U.~S. Army Research Office (under grant W911NF-19-1-0365) and in part by the National Science
Foundation (under grant CCF-1714180). N.~Alipourfard and K.~Lerman are funded in part by DARPA (under contract W911NF1920271) and in part by the Air Force Office for Scientific Research (FA9550-20-1-0224). 
}

\showacknow 


\end{document}